\documentclass{aa}
\usepackage{natbib}
\bibpunct{(}{)}{;}{a}{}{,}  
\usepackage[figuresright]{rotating}
\usepackage{graphicx}
\usepackage{txfonts}
\begin{document}

\title{Variability-selected active galactic nuclei from  supernova search in the Chandra deep field south}


\author{D. Trevese \inst{1}, K. Boutsia \inst{1,2}, F. Vagnetti \inst{3}, E. Cappellaro \inst{4}, S. Puccetti \inst{5} }

   \offprints{Dario Trevese, \email{dario.trevese@roma1.infn.it}}

   \institute
       {Dipartimento di Fisica, Universit\`a di Roma ``La Sapienza'', P.le A. Moro 2, I-00185 Roma, Italy               
         \and
	 European Southern Observatory, Karl-Schwarzschild-Strasse 2, Garching D-85748, Germany
         \and
         Dipartimento di Fisica, Universit\`a di Roma "Tor Vergata'', via delle Ricerca Scientifica, 1, I-00133, Roma, Italy
         \and
	 INAF - Osservatorio Astronomico di Padova, Vicolo dell'Osservatorio 5, I-35122 Padova, Italy 
	 \and
	ASI Science Data Centre, c/o ESRIN, via G. Galilei, I-00044 Frascati, Italy
             }   

 \date{}

  \abstract
{Variability is a property shared by virtually all active galactic nuclei (AGNs), and   was adopted as a criterion for their selection using data from multi epoch surveys.  Low Luminosity AGNs (LLAGNs) are contaminated by the light of their host galaxies, and  cannot therefore be detected by the usual colour techniques.  For this reason, their evolution in cosmic time is poorly known. Consistency  with the evolution derived from X-ray detected samples
has not been clearly established so far, also because the low luminosity population consists of a mixture of different object types.  LLAGNs can be detected by the nuclear optical variability of extended  objects.}
{Several variability surveys have been, or are being,  conducted for the detection of supernovae (SNe).  We propose to re-analyse these SNe data using a variability criterion optimised for AGN detection, to select a new AGN sample and study its properties.}
{We analysed images acquired with the wide field imager at  the 2.2 m ESO/MPI  telescope, in the framework of the STRESS supernova survey.
We selected the AXAF field centred on the Chandra Deep Field South  where, besides the deep X-ray survey, various optical data exist, originating in the EIS and COMBO-17 photometric surveys and the spectroscopic database of GOODS.}
{We obtained a catalogue of 132  variable AGN candidates. Several of the candidates are X-ray sources.
We compare our results with an HST variability study of X-ray and IR detected AGNs,  finding consistent results.
The relatively high fraction of confirmed AGNs in our sample (60\%) allowed us to extract  a list of reliable AGN candidates for  spectroscopic follow-up observations.}
{}


\keywords{Surveys - Galaxies: active - Quasars: general - X-rays: galaxies}
\authorrunning{D. Trevese et al.}
\titlerunning{Variability-selected AGNs}
\maketitle

\section{Introduction}

The optical variability of quasars (QSOs) was discovered even before the very nature of their emission lines was understood \citep{matt63}.
A sharp decline in cosmic time of both radio- and optically-detected QSO populations was established soon after  
their discovery \citep{schm68,schm70}.
To our knowledge, the use of variability as a tool to detect QSOs
was proposed for the first time by \citet{berg73}.
Any of the  QSO properties, such as non-stellar colour, broad emission lines, or variability, can be used to select statistically relevant QSO samples.  The comparison between samples selected using different techniques  enables to
evaluate the relevant selection effects and to derive the intrinsic cosmological evolution of the QSO population. Since the early days of QSO astronomy, we have learned that different techniques can detect different but related classes of objects; this occurred for the UV excess selection technique,  and led to the discovery of radio-quiet QSOs, which are 5-10 times more numerous than radio-loud ones \citep[e.g.][]{kell89,jian07}. The present knowledge of the evolution with cosmic time of the QSO luminosity function (LF) is based mainly   on
the 2QZ survey \citep{boyl00,croo01} for  $z<$ 2.5, on  spectroscopic  surveys for  $z\ge$ 3 
 \citep{warr94,schm95}, and on 
Sloan Digital Sky Survey (SDSS) data for $z>$ 4.5 \citep{fan01,and01}.  
These analyses led to a consensus scenario where  a rapid increase in the QSO number
density with cosmic time , down to $z\sim$2, is followed by a slower decline in the LF, which can be described
by luminosity evolution. This view  was confirmed  by   \citet{rich06}, whose analysis,  however, is not the best suited to locate the epoch at which the number density of active galactic nuclei (AGNs) reached a peak.  Both 2QZ and SDSS QSO candidates 
are selected on the basis of non-stellar colours. 
This selection technique
is limited to point-like objects, i.e. bright  active nuclei outshining the host galaxy,  which, would  otherwise  produce non-stellar colours.  The most accurate determination of the epoch of maximum density
was determined by \citet{wolf03}, on the basis of the COMBO-17 survey, which 
provided a ``low resolution spectrum''  enabling the "point-like" condition to be neglected,  for selecting AGNs on 
basis of their  SED alone. The analysis by \citet{wolf03} indicates that the epoch of maximum density
corresponds to $z \simeq 2$ and  it is independent of QSO luminosity.
Even in this case, however,
the selection is limited to nuclei brighter than $M_B \simeq -21.5$, since otherwise the SED is dominated
by the host galaxy light,  which prevents  the nuclear spectrum from being recognised.
Since \citet{berg73}, variability was adopted as a tool for selecting QSOs/AGNs in various studies \citep{ushe78,hawk83,cris90,t89,ver95,geh03}. An important aspect of variability as an AGN search technique is that it can be applied to extended objects. These include  low luminosity AGNs (LLAGNs) that cannot be detected by the (non-stellar) colour selection
since their SED is contaminated (or even dominated) by the light of the host galaxy.  In this case, variability selection becomes easier since nuclear variability tends to increase as nuclear luminosity decreases  \citep{hook94,t94,cris96}. \citet{btk98} selected a sample of ``variable galaxies", i.e. galaxies with variable nuclei, in the field of Selected Area 57, where other detection
techniques had been applied: these techniques included the use of colours and absence of proper motion  \citep{kkc86,koo88}, and for several objects the AGN character was demonstrated either by spectroscopic observations
\citep{trev08} or by X-ray emission \citep{trev07} (see below).


In many respects, the best way of detecting AGNs is just to use  X-ray surveys, which enable us to distinguish
between stellar  and accretion-powered sources, such as AGNs. Furthermore, hard (e.g. 2-10 KeV) X-rays are less affected by dust 
absorption with respect to other bands. The advent of X-ray imaging surveys with Einstein, ASCA, BeppoSAX, and
then XMM-Newton and Chandra have permitted us to 
improve dramatically our understanding of accretion-powered sources and their
cosmic evolution. These surveys  provided the evidence of a strong luminosity dependence in the
evolution, low luminosity sources (i.e. Seyfert galaxies) peaking at 
significantly later cosmic times than high luminosity sources \citep{hasi03,ueda03,lafr05}, at variance with the results of the COMBO-17 survey \citep{wolf03}, an
the redshift of maximum comoving density ranging from $\sim$1 to $\sim $0.5.
It is obvious that part of the discrepancy is due to the difficulty in selecting Type 2 (absorbed) AGNs using optical techniques, but, even restricting our attention to Type 1 objects, the discrepancy maintains and there is  evidence that
this is due to the incompleteness of optical techniques in selecting LLAGNs \citep{bong07}.

Selection by variability has enabled us  to identify objects of intrinsically low X-ray to optical ratio,
which would be otherwise missed by X-ray surveys \citep{sara03,sara06,trev07}. These objects, in addition to their contribution  to the  LF evolution, provide information on accretion and/or on  star-burst activity.

Repeated observations of the same area of sky enable the detection and study of  various classes of astronomical 
objects, such as variable stars, supernovae, planetary systems, fast moving objects and AGNs.
The Large Synoptic Survey Telescope\footnote{http://www.lsst.org} (LSST) foreseen in 2014 \citep{ivez07} will address most of these issues and will be able to detect AGNs with high completeness to very faint limits
\citep{gree07}. In the mean time, wide field and deep supernova surveys  like ESSENCE \citep{mikn07}, are already providing
data that can be analysed for AGN selection. 
In this paper we report the AGN detection through variability in one of the fields of the 
Southern inTermediate Redshift ESO Supernova Search (STRESS) \citep{bott08}  and we discuss the properties of the variability-selected objects.
The paper is organised as follows. Section 1 describes the data; Sect. 2 describes the method adopted to select the AGN candidates and
the resulting sample; Sect. 3 discusses the optical properties of the AGN  candidates; Sect. 4 discusses their X-ray properties
and Sect. 5 contains a summary of the results.
We adopt throughout the cosmology H= 75 km s$^{-1}$ Mpc$^{-1}$,  $\Omega_M$=0.3, $\Omega_{\Lambda}$=0.7.

\section {The STRESS supernova search programme}
Our detection of variable objects is based on a new analysis of  images from the
Southern inTermediate Redshift ESO Supernova Search (STRESS), which is a long term  project designed to measure the evolution in cosmic time
of the rate of  all types of Supernova (SN) events \citep{capp05,bott08}. The  supernova search is based on the  comparison of images of selected sky fields obtained at different epochs. In general, the temporal sampling of the observations is tuned to the specific goal  to be achieved. To ensure that all SNe are detected within the time that elapsed between the first and the last observation of a given field, the time baseline must be longer than the time for significant luminosity evolution of all SN types, i.e.   as long as 3-4 months. 
For STRESS,  21 fields were initially  selected. They are evenly distributed in right ascension, and have been monitored for about 2 years with an average sampling of one observation every three months. 
A typical observing run was divided in two parts: the search for and  follow-up observation of candidates. The search was conducted during two consecutive nights at the ESO/MPI 2.2 m telescope at ESO, La Silla (Chile). The telescope was equipped with the Wide Field Imager (WFI) and a mosaic of 2x4 CCD detectors of 2048x4096 pixels that image a sky area of ~0.25 deg$^2$ with an excellent spatial resolution of 0.238 arcsec/pixel.
When possible, the first observing night was dedicated to obtaining deep V band exposures for candidate detection, while in the second night the same fields were observed using a different filter, B or R, to collect colour information both for the candidates and the galaxies.  Due to a number of technical, meteorological, and scheduling constraints, in many cases it was impossible to maintain this observing strategy. This implies that only in a few cases
it was not possible to derive the candidate colour. For this reason, in the following we consider only  the V band exposures and derive the candidate colours from another survey, as discussed in Sect. 4. 
To remove detector cosmetic defects, cosmic rays, satellite tracks and fast moving objects,  we acquired for each field three frames
 dithered by 5-10 arcsec, for a total exposure time of 900s or in some cases 600s, as reported in Table \ref{Tab1}.       
For each field,  the difference between the image to be searched (target image) and a suitable archive frame
(template image) was  computed. After accurate astrometric and photometric registrations, the most crucial step in this process was
the matching of the point spread function (PSF) of the two images. This was done using the ISIS2.1 package (Alard 2000)
that, from  comparison of the same sources in the two images, computes a spatially-varying convolution kernel to degrade
the image with the best seeing to match the other one. Variable sources leave residuals in the difference image, which were detected and logged into a catalogue using the  SExtractor program \citep{bert96}. After series of checks to remove false detections, residuals of moving objects and variable stars, one is left with a few SN candidates, typically from none to a handful per field. 
More details on the photometric analysis and the follow up spectroscopic campaign  are described in \citet{capp05} and \citet{bott08}.
For the purpose of the present work, we  note only that, on the basis of spectra obtained at VLT, about 75\% of the supernova candidates
were confirmed and almost all of the remaining 25\% were found to be AGNs.

The approach to candidate selection was designed to avoid as far as possible any selection bias and in particular  nuclear candidates
were  not excluded a priori. Given that, the intrusion of AGNs is unavoidable. For the purpose of SNe searches, contamination by variable AGNs was reduced by looking at the long term variability history of the candidates. Sources showing long-term, erratic variability, were excluded from the list of SN candidates. 
While AGNs represent a contamination of the SNe sample, they are in fact the targets of the present analysis. Thus, we flag as possible SN contaminants
those objects that present a "single flare". Moreover, we are  only interested in nuclear variation. This suggests the use of aperture photometry instead of image subtraction, and  a
statistical approach to variability detection,  described in the following section.
As shown in the next section, the number of AGN candidates exceed SN candidates by about 44 times in our variable source selection, which is optimised for AGN detection down to  $V\sim24$ mag, i.e. we expect a contamination of less than 3 SNe candidates, which is of minor concern.
In principle, we can select AGN candidates in the entire collection of fields monitored by the STRESS programme.
As a first step, we selected the field named AXAF (after the name of the X-ray AXAF satellite subsequently called Chandra), centred
on 03:32:23.7 -27:55:52 (J2000)  and covering $\sim$0.25 deg$^2$.

Table \ref{Tab1} reports the names, dates, and  exposure times of the set of 8 images used in the present analysis, acquired  at the ESO 2.2m  telescope with the Wide Field Imager in the AXAF field.
   
\begin{table}
\begin{center}
\caption{ \label{Tab1} AXAF field observation log}
         \label{Tab1}
\begin{tabular}{lccc}
\hline
Name of Image&Date&Exposure time&seeing\\
&(yyyy-mm-dd)&(s)&(arcsec)\\
\hline
AXAF\_V\_19991109&1999-11-09&900&1.08\\
AXAF\_V\_19991202&1999-12-02&900&1.12\\
AXAF\_V\_19991228&1999-12-28&900&1.15\\
AXAF\_V\_20001116&2000-11-16&600&0.92\\
AXAF\_V\_20001217&2000-12-17&600&0.93\\
AXAF\_V\_20011112&2001-11-12&900&1.03\\
AXAF\_V\_20011118&2001-11-18&900&0.84\\
AXAF\_V\_20011208&2001-12-08&900&0.95\\
\hline
\hline
\end{tabular}
\end{center}
\end{table}

This choice is motivated by the fact that this  field 
overlaps:  i) the ESO Imaging Survey (EIS) \citep{arno01} containing B,V,R, and I photometry and a morphological classification based on 
the {\it SExtractor} code;  ii) the COMBO-17 survey \citep{wolf04}  containing  photometry in 3 broad and 14 narrow bands and
providing a classification of  objects in galaxies, AGNs and stars on the sole basis of their spectral energy distribution (SED); iii) the Chandra Deep Field South (CDFS) survey \citep{giac02,alex03}  that is based on X-ray exposures of 1 Ms, which has spectroscopic follow up observations described  by \citet{szok04}; 
 iv) the Extended Chandra Deep Field South (ECDFS) survey, consisting of four 16.9 x 16.9 arcmin$^2$ pointings of 250 ks each, flanking the CDFS \citep{lehm05}; 
v) the GOODS survey,  which
gathers the data from various catalogues, including the above-mentioned ones, and provides also optical spectra for a sizable fraction of our sample (mostly from \citet{giav04}).  

XMM-Newton observations for a total time of $\sim$500 ks and centred on the CDFS are also available \citep{stre04,dwel06}, but they cover only a fraction of the AXAF field, at a slightly shallower depth with respect to the 1Ms Chandra observation. 
 
\begin{figure}
   \centering
\resizebox{\hsize}{!}{\includegraphics{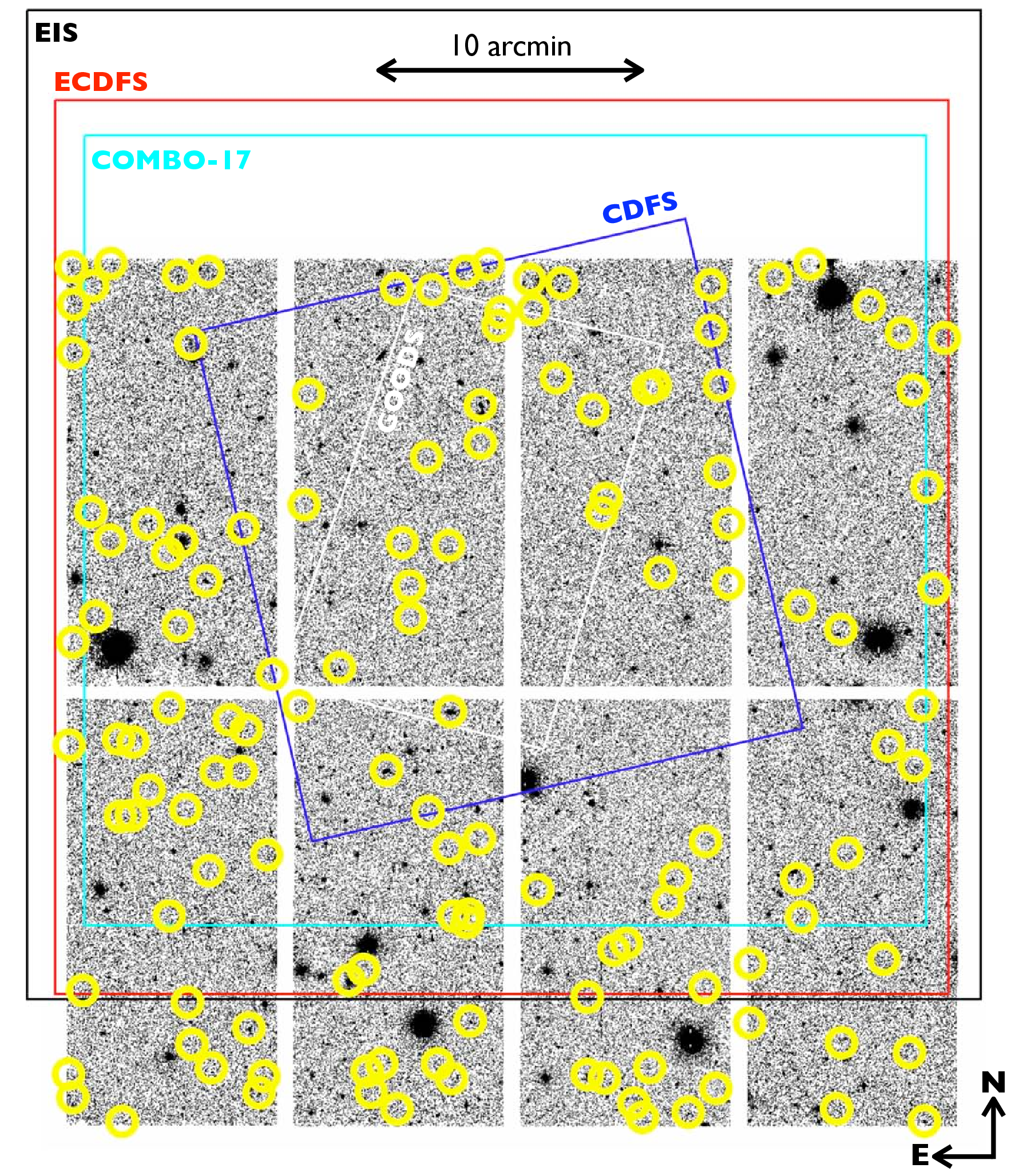}}

      \caption{V band image of the AXAF field taken with the Wide Field Imager at the 2.2m ESO telescope. Contours delimit the area of:  i) the ESO Imaging Survey (EIS) \citep{arno01} ({\it black}); ii) the COMBO-17 survey    \citep{wolf04} ({\it cyan}); iii) the Chandra Deep Field South (CDFS)  1 Ms survey  \citep{giac02,alex03}  ({\it blue}); the Extended Chandra Deep Field South (ECDFS)  250 ks survey  \citep{lehm05}  ({\it red}); v) the GOODS survey ({\it white}) \citep{giav04}. Yellow circles represent the  variable objects reported in Table \ref{Tab2}. 
}
         \label{Fig1}
   \end{figure}

A V-band image of the field is shown in Fig. \ref{Fig1}, where the fields of the above surveys  are also indicated.

In this way, it is possible to analyse the properties of objects selected by means of variability,  by using, in particular, X-ray emission and optical spectra, whenever available;  data about  optical and  image extension also provide valuable information.

\section{Photometry and selection of variable objects}
Objects were detected using {\it SExtractor}   from each of the 8 images corresponding to  different epochs of 
observation of the AXAF field. The image AXAF\_V\_20011208, corresponding to the 8th epoch (see Table \ref{Tab1}), was select to be the  reference image since it contains the highest number of objects, due to its  higher  quality, and has  the best overlap area with the field of the other images. The corresponding catalogue was then assumed to be the reference catalogue.  Aperture photometry  was performed for each object on all images at the same  $(\alpha,\delta)$ positions, for various apertures.

\begin{figure}
   \centering
\resizebox{\hsize}{!}{\includegraphics{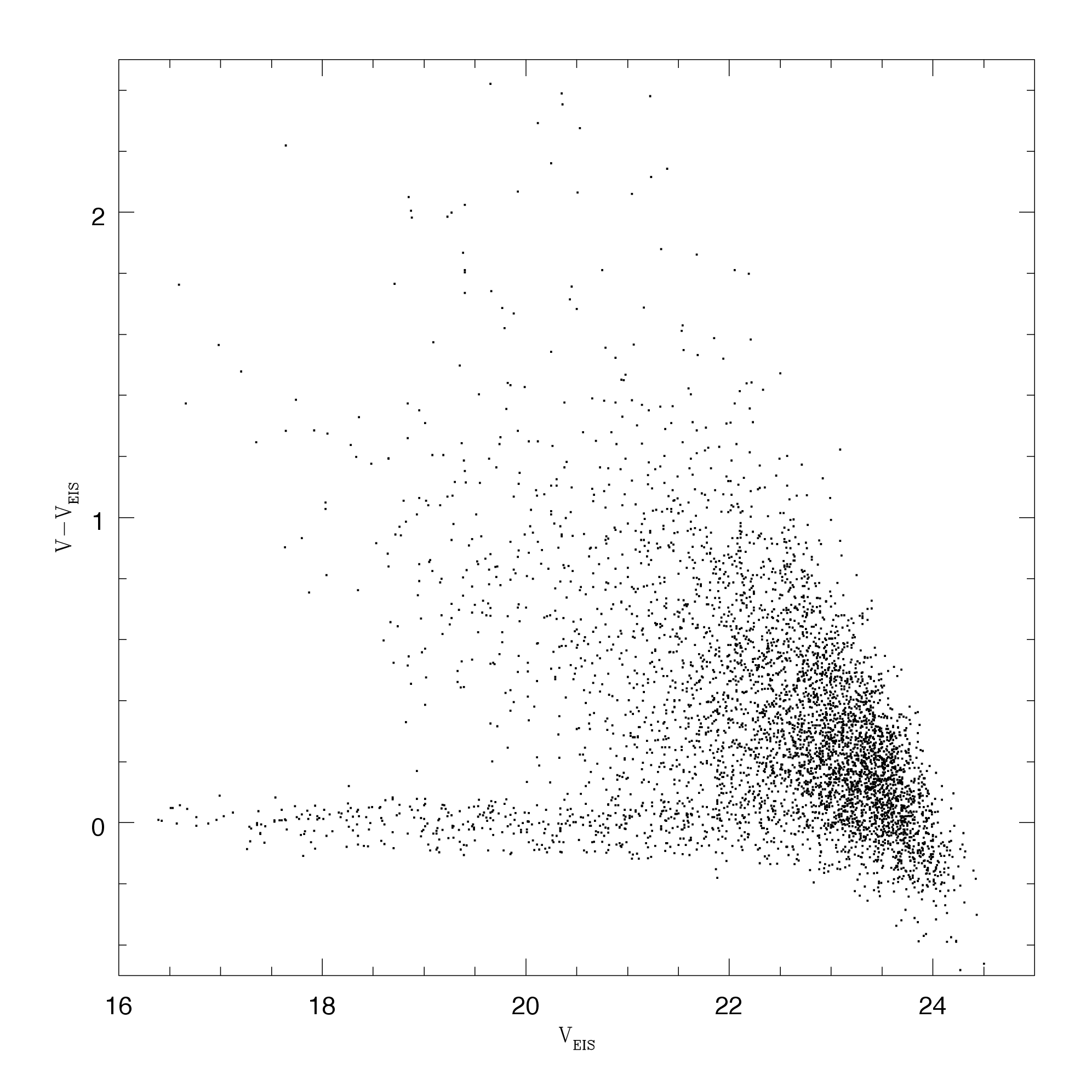}}

      \caption{The difference   $V-V_{EIS}$  between the magnitudes we measured on  2.2m WFI images and those reported in the EIS catalogue. 
}
         \label{Fig2}
   \end{figure}

The V magnitudes, reported in column 4 of Table \ref{Tab2},  are  scaled to  (AB) system magnitudes,
as taken from the EIS catalogue that we use for colour information. More specifically the zero point of our scale is defined by the condition $\langle V-V_{EIS}\rangle$=0, where the brackets indicate the average  taken over all sources with the following constraints: i) non variable, i.e.
 $\sigma^*\leq 3$ (see below) ; ii)   $17<V_{EIS}<21.5$ mag; iii) point-like,  i.e. with {\it SExtractor} stellarity index $>0.9$.
The reason for the latter condition is that we used fixed aperture magnitudes, while  the EIS catalogue reports "total" magnitudes as measured by {\it SExtractor};  for diffuse objects, $V-V_{EIS}$ therefore depends on the extension of the image,  becoming  another indicator of stellarity. 
This can be seen in  Fig. \ref{Fig2}, where stars occupy a "stellar locus" about $V-V_{EIS}=0$, and galaxies are spread over the upper part of the plot.
The r.m.s.  dispersion of the stellar locus provides an indication of the  (quadratically) combined photometric uncertainties of  EIS 
magnitudes  and our own, which  are 0.048,  0.056, and 0.11, for $17<V_{EIS}<19$,  $19<V_{EIS}<21.5$, and $21.5<V_{EIS}<22.5$ respectively. The completeness limit of our catalogue is about $V=24$.

Given any image pair  (i,j), we compute the mean square difference $\langle(m_i-m_j)^2\rangle \equiv \Sigma_{i,j}^2=\Sigma_i^2+\Sigma_j^2$  between the magnitudes of all objects at two epochs (i,j), where the angular brackets represent the average computed over the objects ensemble,
 $\Sigma_i$ and $\Sigma_j$  denote the photometric noise of each image;  we neglect the contribution of intrinsically variable objects,  since they represent a small fraction of  objects in the field. Of course,  photometric noise and  optimal aperture depend on both observing conditions
and  apparent {\it magnitude} of the object, although in reality the  dependence is not extremely strong. For simplicity,  we therefore adopted a fixed aperture  of 
4 pixels (0.92 arcsec) radius,  for all magnitudes and epochs.

Indicating by  $m_k^o$  the magnitude of the k-th object in the reference image, and  by   $m_k^i$  the magnitude of the  same  object in the i-th image, and defining 
$d^{(i,o)}\equiv\langle m_k^i-m_k^o\rangle$ to be
the ensemble average  in a given magnitude bin, we  computed the relative magnitudes
$\mu_k^i(m) \equiv (m_k^i-m_k^o)-d^{(i,o)} $, for each object $k$ and each epoch $i$.  
The amplitude of the magnitude bins was
the result of a trade-off  between the number of objects in each bin and the necessity of tracing the dependence of $d^{(i,o)}$ on $m$.
The  computation of the ensemble average was repeated after a $3$-$\Sigma_{i,j}$ clipping procedure to minimise the effect on $\mu_k^i(m)$
of the most deviant points, which probably correspond to intrinsic variations  rather than noise fluctuations.

\begin{figure}
   \centering
\resizebox{\hsize}{!}{\includegraphics{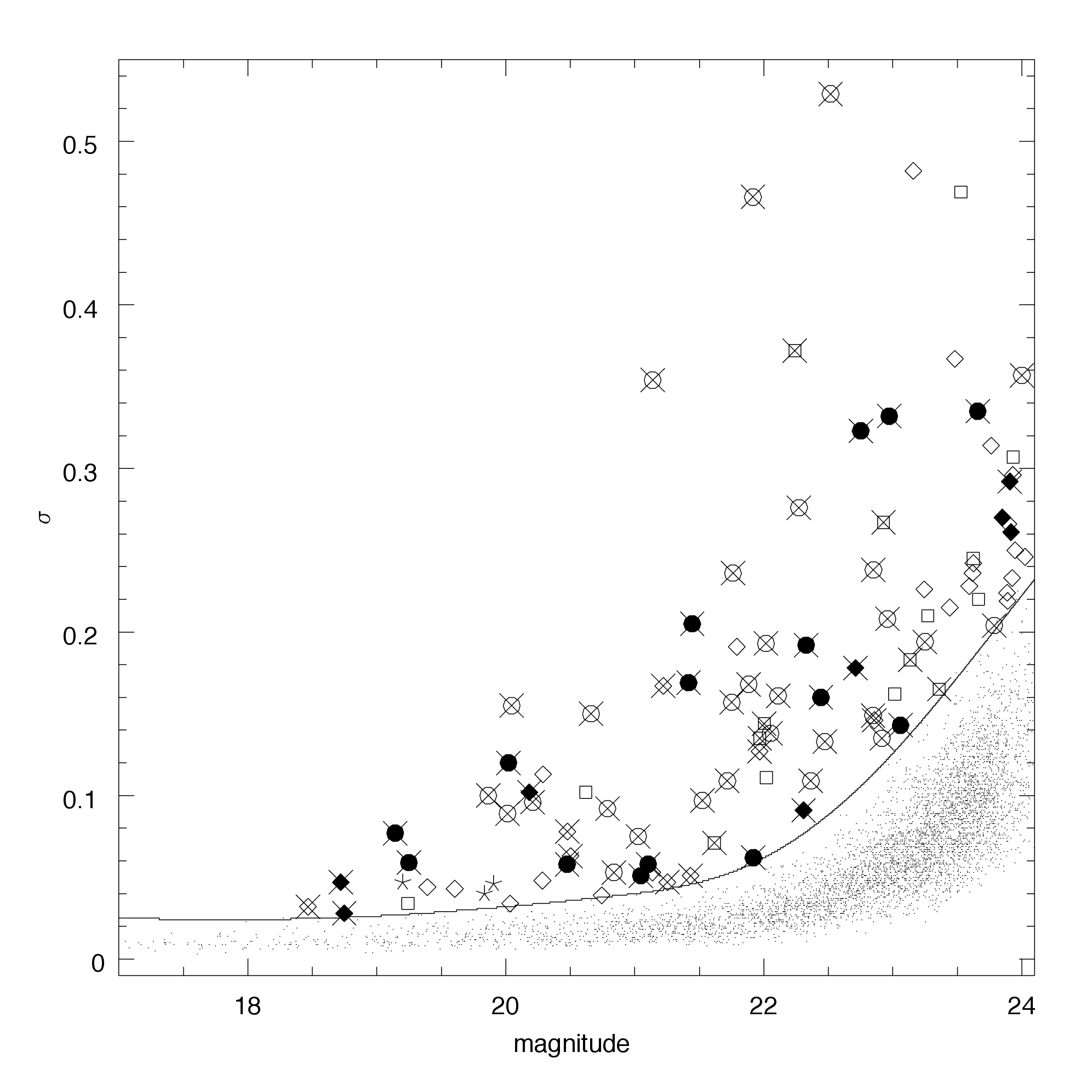}}
 \caption{The r.m.s. magnitude variations $\sigma(V)$ as a function of the apparent $V$ magnitude, for all of the objects in the field. 
      Non variable objects are shown as small dots. Symbols representing the classification are not reported below the  3-$\sigma$ variability threshold, which is indicated by the continuous line.
      Variable objects, above the variability threshold, are indicated as follows.  Filled symbols: objects with spectroscopic redshift; empty symbols: objects without spectroscopic redshift; crosses: X-ray detected objects; circles, diamonds, stars: objects classified by COMBO-17 as QSOs, galaxies, stars, respectively; squares: objects outside the COMBO-17 field.  
     }
         \label{Fig3}
   \end{figure}

In this way, we derived, for each object in the field, relative light curves, which are independent of the observing 
conditions (exposure, background light, seeing, CCD amplification).
From the light curve of each object $k$,  we then compute the mean  $\overline{m}_k$ and the r.m.s. deviation $\sigma_k$ :
\begin{equation}
\bar{m}_k= \frac{1}{N_{epo}} \sum_{i}^{N_{epo}} m_k^i,  \qquad \sigma_k=  [\frac{1}{N_{epo}} \sum_{i}^{N_{epo}} (m_k-\overline{m}_k)^2]^{\frac{1}{2}},
\end{equation}

\noindent where $N_{epo}$ is the number of epochs, which in our case is  8, or smaller in the case of some objects,  close to the
border, which at some epoch are located outside of the field or are contaminated by  hot pixels  or other defects.
In any case, we decided to consider only objects that had measured magnitudes  for at least  5 of the 8 epochs of observation;
in this way, we rejected about 2\% of the objects  and our final reference catalogue contained 7267 objects brighter than $V=24$.
Figure \ref{Fig3}  shows $\sigma \equiv \sigma_k$ (hereinafter we omit the k subscript for simplicity) versus the $V$  magnitude for all  objects in our catalogue. 
The dependence  of both 
the average $s(V)=\langle\sigma\rangle$ and the
r.m.s. deviation $\Sigma_{\sigma}(V)=\langle[\sigma-s(V)]^2\rangle^{1/2}$ on $V$,  where the average is computed in  bins of V, is clearly evident.
The average $s(V)$ represents the r.m.s. noise that must be subtracted from $\sigma$ to measure the intrinsic variability.
As the number $N_{epo}$ of observations is increased, $s(V)$ does not vary, but instead it converges to the r.m.s. noise.
The $\sigma$ of intrinsically variable objects
may or may not change, depending on how the intrinsic variability timescale compares with the total duration of the observing campaign and the 
sampling intervals. The r.m.s. deviation $\Sigma_{\sigma}(V)$, instead, decreases as $N_{epo}^{- 1/2}$, and thus an increase of the number of
observations  allows us to decrease the threshold which defines  variable objects.
To maintain the number of spurious variables produced by the photometric noise at a sufficiently low level, we adopt a  3-$\Sigma_{\sigma}$ threshold:
\begin{equation}
\sigma\ge [s(V)+ 3 \Sigma_{\sigma}(V)].
\end{equation}
The continuous line  in Fig. \ref{Fig3} represents the adopted threshold. The r.m.s. deviation
 $\sigma$ is a measure of the average amplitude of magnitude changes,  which are due in part  to photometric noise and in part to intrinsic variability. 
We can also define for each object a normalized r.m.s. deviation (as in \citet{btk98}):
\begin{equation}
\sigma^*\equiv \frac{\sigma- s(V)}{ \Sigma_{\sigma}(V)},
\end{equation}
which provides a measure of the significance of the variability. Thus variable objects are defined according to the condition $\sigma^*\ge3$.

From the above discussion, it is clear that this method is not optimal for supernova detection. For instance, if the sampling 
interval between two observations is comparable or larger than the timescale of SN decay, then $\sigma$ {\it decreases} with the number $N_{epo}$ of observing epochs.
For AGNs, instead, $\sigma$ increases on average, at least for delays up to $\sim$ 50 years, as shown by the structure function analysis of AGN variability \citep{devr03,devr05}.

Our procedure, once applied to the entire AXAF  field, produces a list of  132 candidates reported in Table \ref{Tab2}.
A spectroscopic follow-up project is necessary in any case,
not only to confirm the AGN nature of the candidates and exclude other kinds of variable objects, but also to measure their  redshift  and  assign them to  specific classes, such as type 1 or 2 Seyfert galaxies, QSOs,  low luminosity AGNs, or star-burst galaxies. 

The threshold applied was chosen 
to be a trade off between the completeness level that we want to achieve and the fraction of spurious candidates  (purity or reliability) that we are willing to accept
in a follow-up spectroscopic campaign.  Purity and completeness are discussed in Sects. 4 and 5 respectively.

\section{ Variability-selected AGN candidates}

\begin{figure}
   \hspace{-2.7cm}
    \includegraphics[width=12cm]{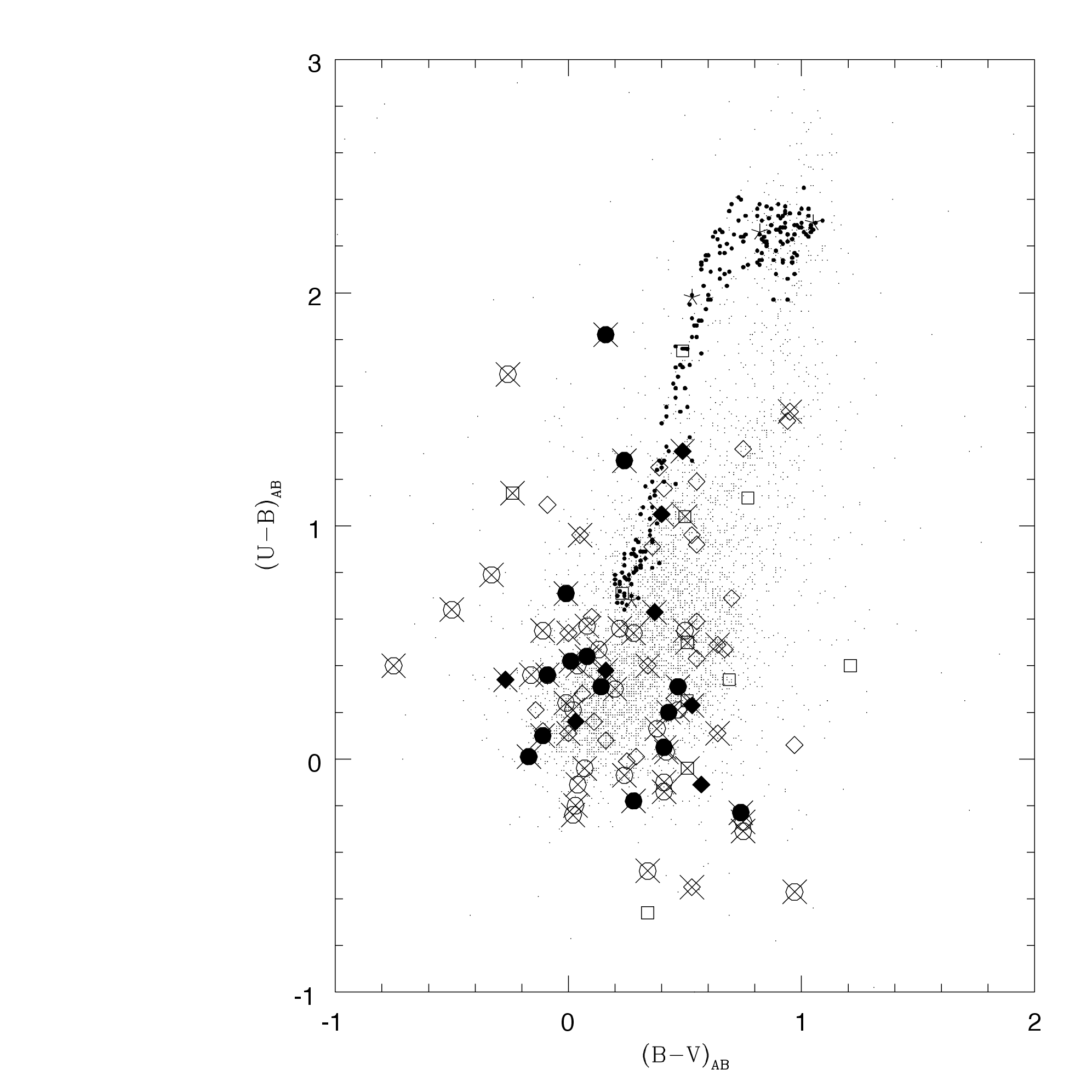}
      \caption{(U-B)$_{AB}$ versus (B-V)$_{AB}$ colours for all 5138 objects with EIS photometry in the AXAF field and measured  at least at 5 epochs. Non-variable galaxies: small dots; non-variable stars: larger dots. Variable objects are indicated as follows: filled symbols, objects with spectroscopic redshift; empty symbols, objects without spectroscopic redshift; crosses, X-ray detected objects; circles, diamonds, stars: objects classified by COMBO-17 as QSOs, galaxies, and stars, respectively; squares, objects outside the COMBO-17 field.
      }
         \label{Fig4}
   \end{figure}

\begin{figure}
   \centering
\resizebox{\hsize}{!}{\includegraphics{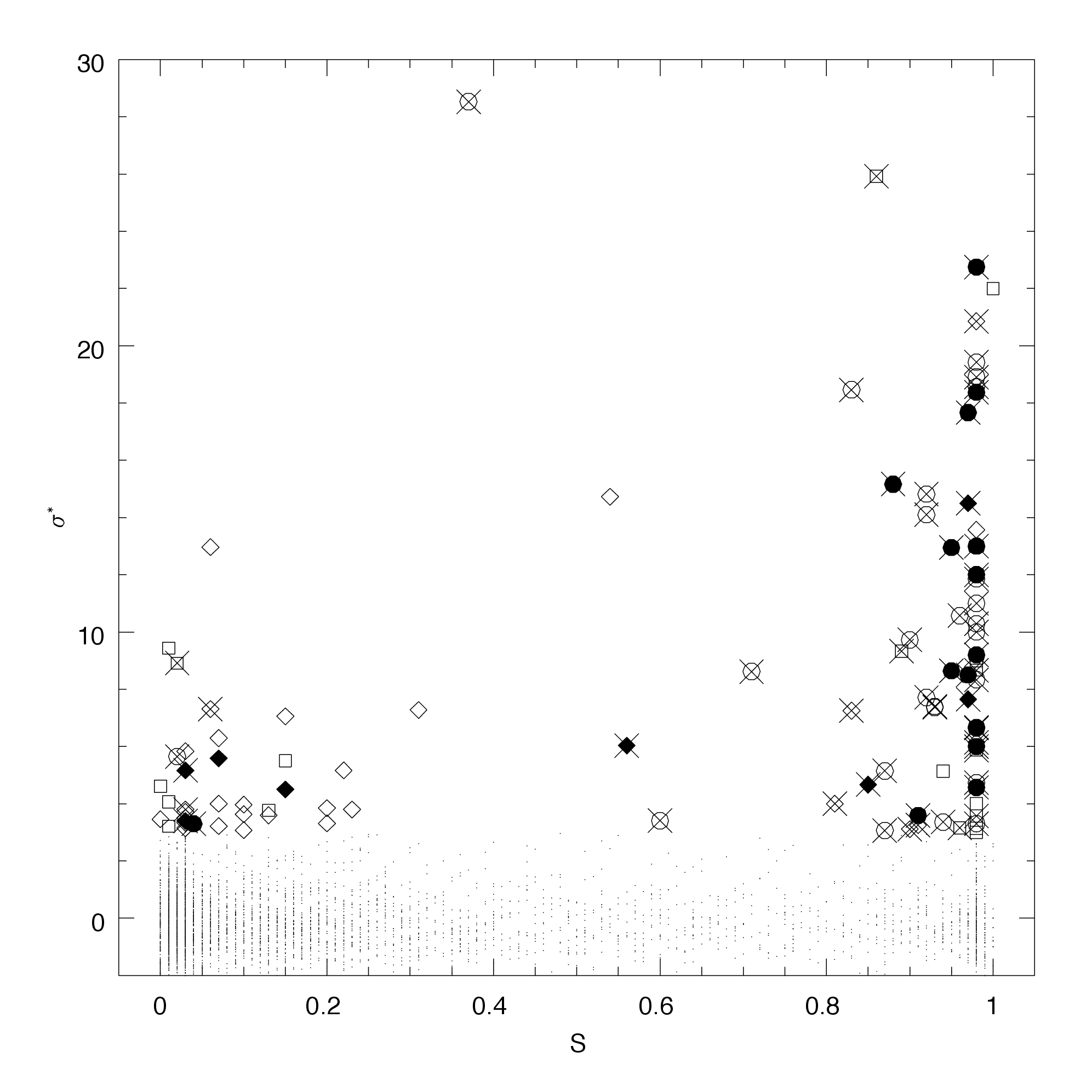}}

      \caption{Normalised r.m.s. variability $\sigma^*$ versus the {\it SExtractor} stellarity index $S$.
      Symbols as in Fig. \ref{Fig3} and  \ref{Fig4}.
}
         \label{Fig5}
   \end{figure}

The colour-colour diagram of  Fig. \ref{Fig4} shows all 5138 objects  with EIS photometry in the AXAF field, measured  at  5 epochs at least.
Large symbols represent variable objects reported in Table \ref{Tab2}, possessing EIS photometry in the relevant bands:  all of them (but one) belong to the ECDFS area (see Fig. \ref{Fig1}). 
Since the fraction of area covered by EIS and not by COMBO-17 is small (see Fig. \ref{Fig1}), most objects possess  a COMBO-17 classification. More specifically, only 10 objects, represented as open squares, do not possess this classification:  4 of them  are detected in X-rays (crosses). 
Although spectroscopy is required to measure their redshifts and to achieve more detailed classifications, these 4 objects can be considered to be AGNs on the
basis of the presence of both variability and X-ray emission \citep[see][]{maoz05}.
We evaluate, in addition,  their X-ray to optical ratio (X/O), that is defined on the basis of  R-band optical flux and 2-8 keV X-ray flux, which is less affected by obscuration than softer X-ray bands. All of these objects have
 $\log X/O > -1$, a value more consistent with AGN activity  than  starburst galaxies, which typically have $\log X/O< -1$ \citep[e.g.][]{geor07}. Obviously the opposite is not true, since the host galaxy light may reduce the apparent X/O ratio of a faint AGN, depending on the aperture adopted for optical photometry and the seeing conditions.
The other objects represented by open symbols are either QSOs (circles) or galaxies (diamonds), with COMBO-17 classification but without a optical spectroscopy.
All of the former are also X-ray emitting and,
if we consider the high  reliability of the COMBO-17 classification for objects dominated by  nuclear emission, 
we can assume that they are  {\it bona fide} QSOs.  Among the variable objects classified as galaxies by COMBO-17 and without spectroscopic redshift (open diamonds),  some (7) have also X-ray emission and are likely to be relatively faint AGNs hosted by a galaxy whose light swamps the nuclear radiation.
This is indicated  by their stellarity index in Fig. \ref{Fig5} where most of them (6/7, open diamonds with crosses) are non stellar. The same can be true for  objects not detected in X-rays, although, in this case,
the nuclear component  must be even fainter. Their variability suggests that they are AGN, which, however,  requires  spectroscopic confirmation. If the redshift is known, an X-ray luminosity
$L_X($2-8 keV$)> 10^{42}$ erg s$^{-1}$ can be assumed as a more direct indication of the AGN character. 
All of the spectroscopically confirmed QSOs (14 filled circles) have X-ray emission. Of the  galaxies with known redshift (8 filled diamonds),
4  have spectra with broad emission lines,   are also detected in X-rays, and are consistent with AGN activity.  
To evaluate the purity of the sample, i.e. the fraction of AGNs in the variability selected sample, we consider the 104 variable objects in the field covered by X-ray data, 4 of which are variable stars (with a variability in the range 0.03-0.05 mag r.m.s). Among these 104 variable objects,  we consider  bona fide AGNs to be: 44 COMBO-17 QSOs with X-ray emission,
4 Combo-17 galaxies with broad emission lines, 8 COMBO-17 galaxies without spectrum but with $\log X/O>-1$, and 7 objects without COMBO-17 classification but with $\log X/O >-1$. This corresponds to a purity of at least 60\% (63/104). We stress that this is a  lower limit, since most of the remaining variable objects could also hide some, possibly low luminosity, AGN component that future spectroscopy could  reveal.   

In Fig. \ref{Fig5}, the normalised r.m.s. variability $\sigma^*$ is reported versus the {\it SExtractor}  stellarity index that  ranges from 0 (extended objects)  to 1 (pointlike objects).   
Small dots are  non-variable sources ($\sigma^* \leq 3$).  Most QSOs, either  spectroscopically confirmed or not, tend to have high values of  stellarity index,  but sometimes between 0.6 and 0.8 and not necessarily very close to 1.0. This indicates that often the "fuzz" due to the host galaxy is detected by the stellarity index.  For two of the COMBO-17 QSOs, one of which is spectroscopically confirmed, the stellarity index is however  typical of  that of a galaxy. This circumstance implies that  further investigation   
is necessary. One possible explanation is related to variability: for example, the nucleus could have been fainter at the epoch of morphological classification and brighter at the epoch of COMBO-17 SED classification. The same is true for objects classified as COMBO-17 galaxies that have a stellarity index close to 1.
It appears that all  4 galaxies with optical spectroscopy  (filled diamonds) and  a stellarity index greater than 0.5  are X-ray emitting and spectroscopically confirmed AGNs. 
 
%
%



\subsection{Comparison with an HST variability survey}

A variability survey with Hubble Space Telescope (HST) Advanced Camera for Surveys (ACS) was undertaken by
\citet{kles07} in the GOODS-South field, which is contained in our AXAF field.
The variability of a composite sample, consisting  of 22 mid-IR power-law sources from \citet{alon06} and 104 X-ray sources from \citet{alex03}, was analysed on the basis of  ACS exposures at 5 different epochs separated by 45 day intervals. The ACS analysis was motivated by the special interest of detecting and studying  LLAGNs.  With the high angular resolution of HST, it is possible to perform photometry using an extremely small aperture: 0.075$"$ radius in this case. This  reduces the diluting effect of the host galaxy light on the nuclear variability, allowing for its detection down to a nuclear magnitude $M_B \sim -15$,  as discussed by \citet{sara03}.  \citet{kles07} defined a variability measure  by the standard deviation $\sigma$ of the light curve, as in the present work, and defined  the error in the variability measure $\sigma$ as 
${\rm error}_{\sigma}=\sqrt{\Sigma({\rm error}_{mag})^2/N}$, where N is the number of epochs in which the object was observed and
${\rm error}_{mag}$ is the formal photometric error in the magnitude at each epoch. Although not identical, the definition of ${\rm error}_{mag}$ is almost equivalent to our empirical estimate $\Sigma_{\sigma}$ based on the magnitude spread of non-variable objects in each magnitude bin.
The  significance of variability is then defined to be: Significance $=\sigma/{\rm error}_{\sigma}$. Thus, under the assumption that error$_{\sigma}$ and our $\Sigma_{\sigma}$ are equal to each other (see Eq. (3)),  the relation Significance $=\sqrt{{(\sigma^*)}^2+1}$ would hold.
We note, however, that the level of noise  is not the same in our WFI observations and  the ACS observations of \citet{kles07}.

\begin{figure}
   \centering
\resizebox{\hsize}{!}{\includegraphics{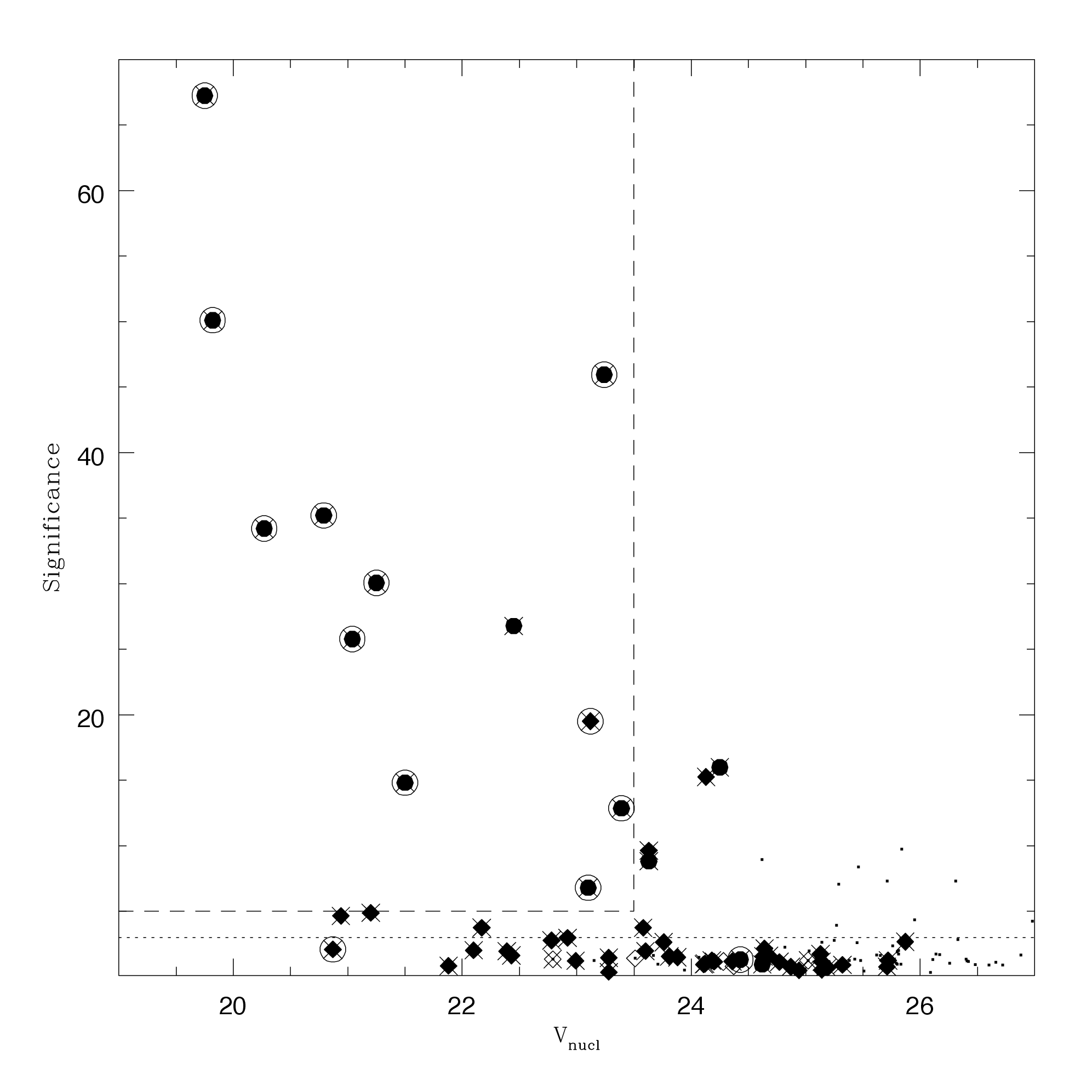}}

      \caption{Significance versus the nuclear magnitude V$_{Nucl}$ from   \citet{kles07}. Small dots: objects not in our survey; large symbols: objects which appear in the present survey; filled symbols: objects with  spectroscopic; open symbols: objects without spectroscopy; diamonds: galaxies  form COMBO-17; circles: COMBO-17 QSOs;  
crosses: X-ray sources; large open circles:   variable objects from our survey. Two of the latter are not variable according to \citet{kles07} (see text). The horizontal dotted line is the variability threshold of  \citet{kles07}. Within the area delimited by the dashed lines, all objects but one are variable according to both surveys.  
}
         \label{Fig6}
   \end{figure}

We  compare our data with ACS variability observations, for all  objects in common. For these objects,  Fig. \ref{Fig6} displays the Significance versus  the nuclear  magnitude $V_{nucl}$ taken from Table 1 of \citet{kles07}. 

Objects that appear  variable in our survey, i.e. for which $\sigma^*\ge 3$, are marked by large open circles.
For $V_{nucl}<23.5$,  11/15 (73\%) objects that are variable according to \citet{kles07}, i.e. with Significance $>3$, are variable in our survey. This fraction increases to 11/12 (92\%), if we consider only objects with Significance $>5$. 

Both limits depend on  photometric noise. A nuclear magnitude $V_{nucl}=23.5$
corresponds, on average,  to a value  of about 23 in our  aperture magnitude scale, where the
3-$\sigma$ threshold becomes of the order of 0.1 mag  (see Fig. \ref{Fig3}).  Moreover, the lower angular resolution of ground-based observations
tends to dilute nuclear variations and increase the flux-variability threshold, for a given significance.
In Fig. \ref{Fig6}, there are two objects that appear variable in our survey but lie below the Significance threshold defined by \citet{kles07},  which is given by the dotted line in Fig. \ref{Fig6}. One of these objects is an X-ray detected and spectroscopically confirmed QSO with $V_{nucl}=24.43$  (ID~115 in Table\ref{Tab2}). The second (ID~5) is a  galaxy  with an extra-nuclear Ultra-Luminous X-ray source (ULX) \citep{horn04,lehm06}, which could produce extranuclear variability, missed by the small-aperture nuclear photometry of \citet{kles07}. In any case, the WFI and ACS observations  were performed in different periods,  and for a total sampling time of 0.5 years in the case of  \citet{kles07} and, in our case,  2 years.  It is, therefore,  unsurprising that a few objects are detected as variable in one survey and not in the other. 



\subsection{X-ray properties}

The X-ray emission is a specific signature of AGNs. The  CDFS 1 Ms  \citep{giac02} is one of the deepest X-ray exposures available in the sky.  Additional  1 Ms observations of the CDFS have been performed recently\footnote{http://cxc.harvard.edu/cda/whatsnew.html\#cdfs2000-2007}
and a further analysis of the field, including these data and forthcoming new optical spectra, is in progress \citep{bout08a}.
The AXAF field is therefore ideal for a combined study of the X-ray and optical properties of AGNs. 
In the following, we define X/O to be the ratio of the observed fluxes in the R optical band to the fluxes in the 2-8 keV X-ray band. We neglect the k-correction that, however, acts in the same sense for both the optical and the X-ray band. Therefore we  neglect  a difference in k-corrections,
which is relatively small with respect to the intrinsic spread of the X/O values.   
   
\begin{figure}
   \centering
\resizebox{\hsize}{!}{\includegraphics{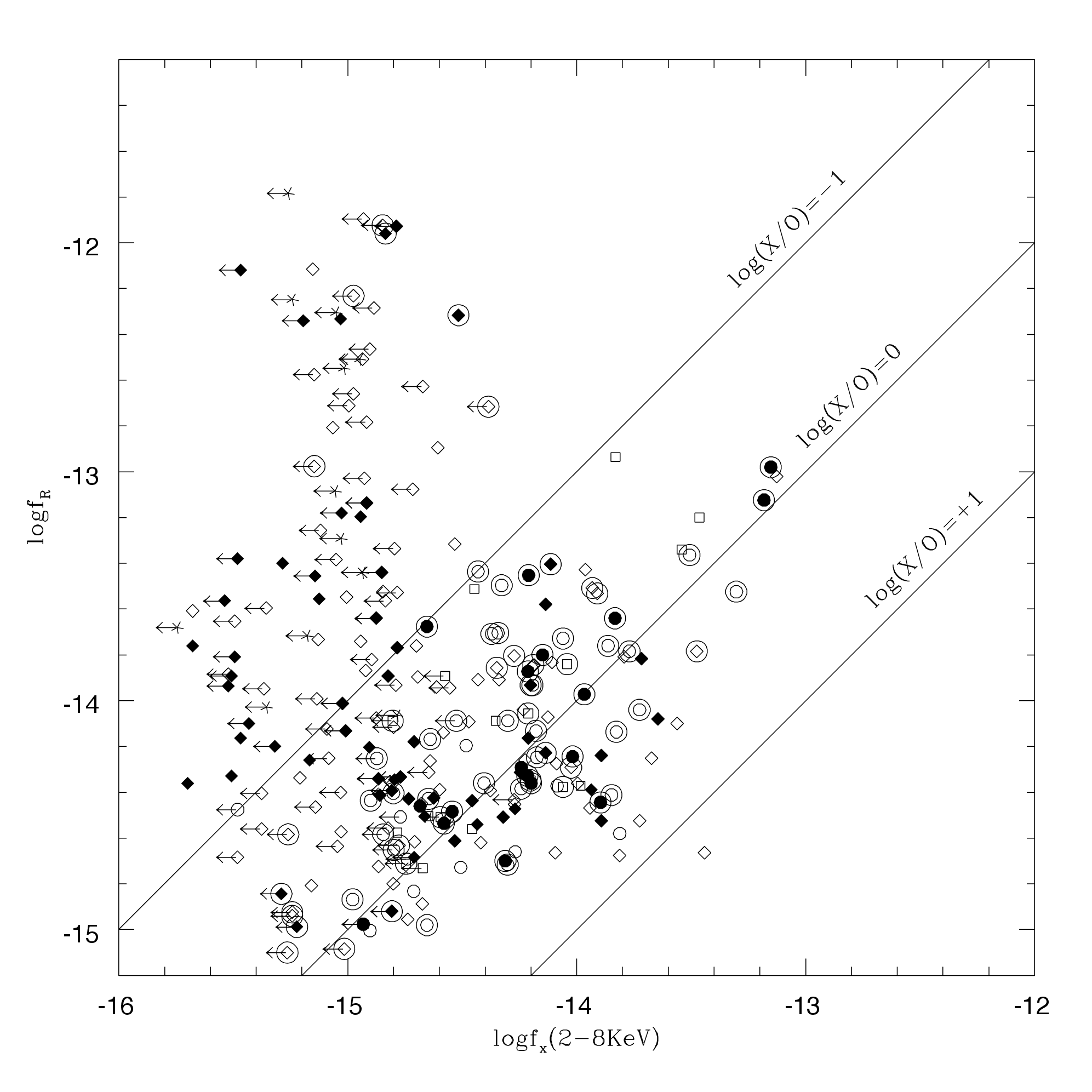}}

\caption{$\log f_R$ versus $ \log f_X(2-8\,{\rm keV})$ for all the objects with measured X-ray flux or with estimated upper limit. Fluxes are taken from \citet{lehm05} when available or from \citet{alex03} or \citet{giac02} otherwise.  Symbols as in Fig. \ref{Fig3}. Variable objects are marked by large open circles.  Arrows indicate available 3-$\sigma$   upper limits from the same surveys, or from our estimates,  for objects undetected   in the $(2-8\,{\rm keV})$band.
}
         \label{Fig7}
   \end{figure}

Due to the very low limiting flux,  of about  $3 \times 10^{-16}$ erg cm$^{-2}$ s$^{-1}$ in the 2-8 keV band, the distribution of the optical versus X-ray fluxes, shown in Fig. \ref{Fig7}, differs from 
 previous studies with brighter X-ray limits. Below $\sim 5 \times 10^{-15}$ erg cm$^{-2}$ s$^{-1}$, a population of very low X/O objects appears.
This is due to the X-ray emission of normal galaxies. Note that 12 objects, classified as COMBO-17 stars, are in the same X-ray flux range and have $ -4\la \log(X/O)\la -1$.

The majority of variable objects (large circles) are concentrated in the $-1<log(X/O)<+1$ stripe in Fig. \ref{Fig7}, which correspond to typical AGNs. However, six variable objects have lower X/O values, all of which have  COMBO-17  SEDs of normal galaxies (diamonds). Two objects (ID 4 and 5 in Table \ref{Tab2}) are detected in X-rays, have low log(X/O) (-2.2 and -2.9), and correspond to extranuclear X-ray sources \citep{horn04,lehm06}. For all the X-ray undetected variable objects, we have calculated 3-$\sigma$ upper limits to the X-ray flux, from the 1Ms Chandra images, if not already available from \citet{lehm05} or \citet{alex03}.  All the upper limits, for variable and non-variable objects, are indicated by leftward arrows in Fig. \ref{Fig7}. The lowest upper limit to log(X/O) for variable objects is equal to approximately -3 and corresponds to the object ID~3.


\begin{figure}
   \centering
\resizebox{\hsize}{!}{\includegraphics{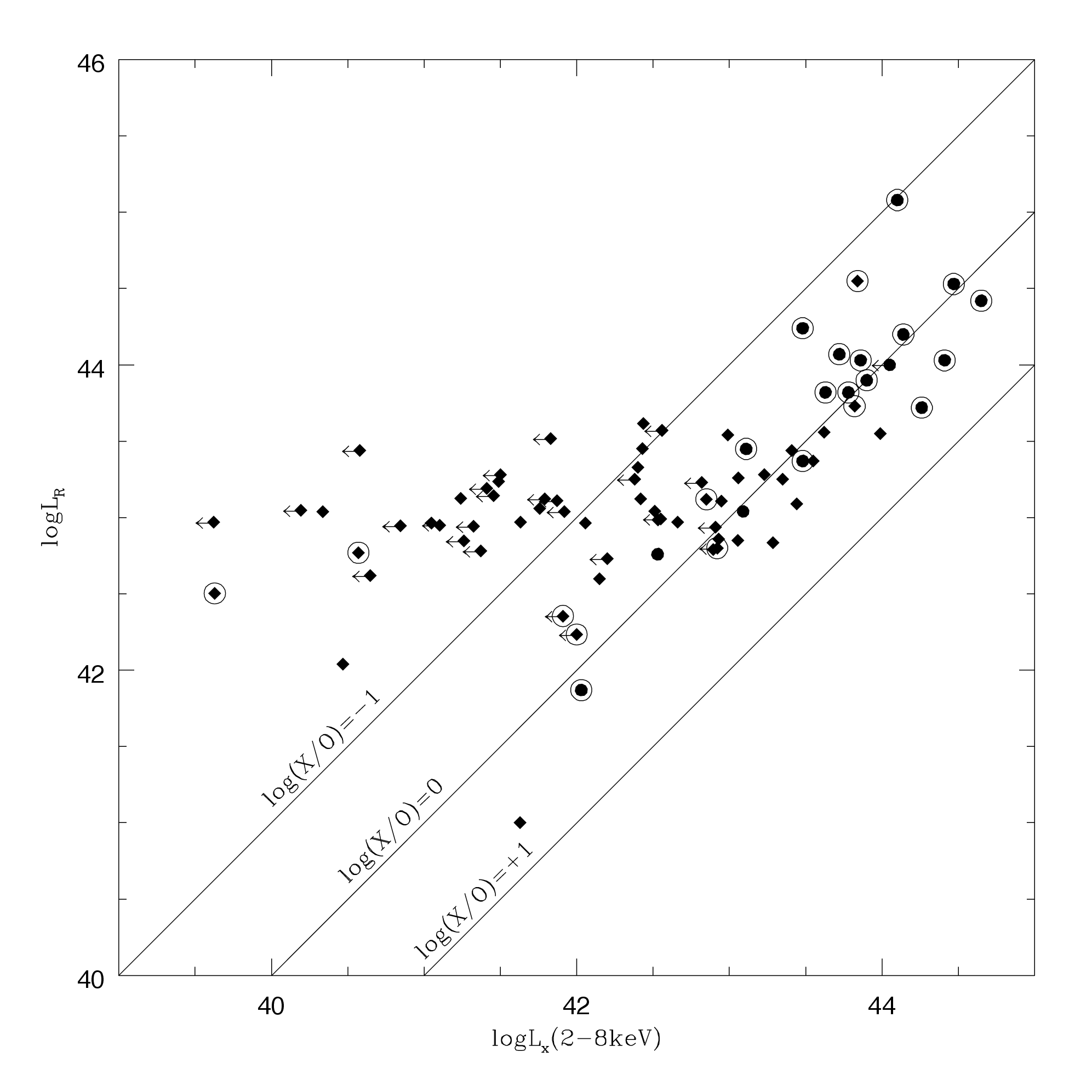}}

      \caption{$\log L_R$ versus $\log L_X(2-8\,{\rm keV})$ for the objects with known redshift. Symbols as in Fig. \ref{Fig7}.
}
         \label{Fig8}
   \end{figure}

For  objects of known redshift, we consider  the optical versus X-ray luminosity distribution, shown in Fig. \ref{Fig8}.  There are 18 objects that are classified as COMBO-17 QSOs, and all 
are broad line AGNs, according to \citet{szok04}, and most of them (15/18) are variable. 
Including COMBO-17 galaxies, there are 23 variable objects, the majority of  which
(19/23) have a measured X-ray luminosity $L_X($2-8 keV$)>10^{42}$ erg s$^{-1}$, a value typically used to distinguish between AGN and normal galactic emission. Most of the variable objects occupy 
the stripe  $-1<\log(X/O)<+1$. We note that 4 of these (21\%) were classified as COMBO-17 galaxies
but show an emission line spectrum of broad line AGNs (ID 23, 75, 83, 125). 
Two (ID 75, 125) have relatively low {\it SExtractor} stellarity index, and would therefore not be detected by usual colour  technique, probably due to the relevance of the host galaxy light.  
Of the  4 variable objects below $10^{42}$ erg s$^{-1}$,  2 are the above mentioned ULXs, and the other 2 have upper limits consistent with normal AGNs.


\section{Discussion and summary}

As discussed in the Introduction,  the discovery of AGN variability precedes  historically even the notion AGN, and various samples have been created on the basis of this selection technique. However the amplitude of variability depends  on the timescale, type of object, and its luminosity. The properties of the selected sample, such as completeness,  purity, and the type of detected objects, depend on the sampling interval,  number of observing  epochs, total duration of the observing campaign, observing band, and photometric accuracy.  The principal aim of the present work was to analyse the properties of AGN candidates selected on the basis of their variability 
from a photometric campaign, whose sampling rate and duration were developed especially for the detection of supernovae. 
Large SNe surveys are (and will be) conducted in view of their crucial importance  primarily in constraining the dark matter/energy of cosmological models, but also  the  evolution in cosmic time of the galaxy population. The creation of large statistical samples of AGNs as a by-product of SNe surveys
may add scientific value to these surveys without increasing their cost in terms of observing time or special requirements.
To demonstrate this, we have selected from the STRESS  survey a field that was observed 8 times over 2 years, and was  studied in detail by various other surveys, including the 1 Ms X-ray survey of the CDFS.
As discussed in Sect. 4, our 3-$\sigma$ variability selection produces a sample of good purity, for which the percentage of "true" AGNs is about 60\% of the total number of AGN candidates. This is a lower limit since a fraction of unconfirmed candidates may also be AGNs.
The completeness of the present sample, estimated considering only objects with known  spectra and  X-ray luminosity $ L_X(2-8)keV \ge 10^{42}$ erg s$^{-1}$, is $\sim$44\% (19/43, see Fig. \ref{Fig8}),
 which may increase with further spectroscopic observations \citep{bout08a}. A relatively high incompleteness suggests caution  in basing evolutionary studies
on a single selection technique.  Still, variability selection may complement the most common
surveys  based on optical colours or X-ray emission, which suffer  different biases. 

We stress that by increasing the photometric accuracy and sampling rate, variability surveys can reach far higher completeness. For instance,
\citet{sesa07} derived $\sim$90\% completeness with a threshold of 0.03 mag r.m.s for objects brighter than g=19.5, maintaining the contamination of the sample at an acceptable level.
A survey to be conducted by the forthcoming LSST may reach  a completeness level of 100\% 
at a limiting magnitude $i < 24$, with 12 exposures distributed over a year, according to numerical simulations \citep{gree07}  based on the extrapolation of statistical properties of AGN variability,  as quantified by \citet{vand04} from a variability study conducted on a sample of  ~25000 QSOs observed by the SDSS.
We notice also that, although variability is one of the principal characteristics of active galactic nuclei, its nature is still poorly understood. Independent models have been suggested, including supernova explosions, microlensing, and accretion disk instabilities \citep{aret97,hawk93,kawa98,trev02}. A comparison of these models is discussed in \citet{hawk07}. Multi-band variability studies of  statistical AGN samples can further constrain the physical origin of luminosity changes.
 
In general, objects that appear to be galaxies, due to their extended images or their SEDs, are lost by the colour technique (see Fig. \ref{Fig4}). This happens in the case of  LLAGNs,  swamped by the host galaxy light. To reach the lowest possible luminosity limit,  variability must be studied using the highest possible spatial resolution, as achieved for HST images both in the HST-N  and in the Groth strip fields by \citet{sara03,sara06}, to reduce the dilution of the nuclear light.  Atmospheric seeing obviously prevents us from reaching the same luminosity limits with ground-based observations.
The comparison of the results for our different data sets, based on $\sim$15 min exposures at the ESO 2.2 m telescope, and the variability study of  \citet{kles07}, based on ACS/HST images of X-ray selected AGNs, indicates that we detect variability at a confidence level of 3-$\sigma$, down to $V_{nucl} \sim 23.5$, for 70\% of the objects detected as variable sources at a confidence level of 3-$\sigma$ by these authors.
Among these objects, the faintest that we observe to be 3-$\sigma$ variable has $V_{nucl}$=24.43, while the analysis of  \citet{kles07} detects variability down to $V_{nucl}\sim27$.
However the area covered by a single field of the STRESS survey is about 900 arcmin$^2$. The STRESS project covers  16 fields, thus can provide
about 16 times the 130 variable candidates of the present study, i.e. ~2000, of which about 1200 are expected
to be {\it bona fide} AGNs on the basis of the high purity (~60\%) of our sample.
This  number would allow us  to assess statistically the selection effects by comparison with colour-selected or X-ray-selected samples.
In a previous paper \citep{trev08}, we analysed the X-ray and optical properties of variability-selected objects in SA 57 and found that several "variable galaxies" are NELGs, for which it was impossible to assess the starburst or LINER nature due to insufficient wavelength coverage. These variable objects are in all cases interesting.
If they are LINERs, their variability would imply that they are AGNs, following
the argument of \citet{maoz05}, 
that the observed variability cannot be accounted for by
the luminosity variations of B stars. We note that this is  {\it a fortiori} true if we consider observations at longer wavelengths than those in the UV 
discussed by \citet{maoz05}, despite the shallow decrease in nuclear variability with redshift \citep{gtv91,tkb91,vand04}. 
If our objects are instead starburst galaxies, the origin of their variability would present an interesting problem, as in the case of the objects ID 4 and ID 5 in Table \ref{Tab2}, which show evidence of extranuclear emission possibly related to ULXs \citep{horn04,lehm06}.

We note that, from data collected so far by the ESSENCE supernova survey, in 32 fields of a total area of 8 deg$^2$,  it should be possible to select about 4,000 candidates on the basis of  variability
\citep{bout08b}; this will increase the number of variable NELGs available and improve our knowledge of the relation between the starburst and AGN phenomenon, providing a development of the synergic SN/AGN survey \citep[cf. conference proceedings about this subject,][]{lampe03,buda04}.

Finally, the ~520 arcmin$^2$ field of the Large Binocular Camera  \citep{gial08} for both red and blue arm of the Large Binocular Telescope,  which has at present the widest field of view available for an 8 m  class telescope,  will provide an unprecedented opportunity for  deep synergic SN/AGN surveys. 


\begin{acknowledgements}
We acknowledge B. Leibundgut  and R. Nesci for discussions and M.T. Botticella for providing details on the STRESS images and help in their handling.
This work was partly supported by MIUR under grant PRIN 2006/025203. K.B. acknowledges support by ESO studentship program. This research has made use of the NASA/IPAC Extragalactic Database (NED) which is operated by the Jet Propulsion Laboratory, California Institute of Technology, under contract with the National Aeronautics and Space Administration.
\end{acknowledgements}

%
%
%
\pagestyle{empty}
\begin{sidewaystable*}
\begin{minipage}[t][180mm]{\textwidth}
\caption{Catalogue of the variable objects}
         \label{Tab2}
\begin{tabular}{rccccrccccccrcccrr}
\hline\hline             
ID&$\alpha$(J2000)&$\delta$(J2000)&$V$&$\sigma$&$\sigma^*$&EIS-name&$U_{EIS}$&$B_{EIS}$&$V_{EIS}$&$R_{EIS}$&$S_{EIS}$&COMBO&class$^a$&$z$&notes$^b$&XID$^c$&$f_X$(2-8keV)$^c$\\
(1)&(2)&(3)&(4)&(5)&(6)&(7)&(8)&(9)&(10)&(11)&(12)&(13)&(14)&(15)&(16)&(17)&(18)\\
\hline
  1& 03:33:19.00& -27:41:44.9&16.97& 0.031& 22.0& J033319.00-274145.2&19.53 & 17.55&  17.02 & 16.51&  1.00 & 47439& S  &      &          &   &           \\
  2& 03:33:09.36& -28:07:25.7&17.13& 0.035&  4.6& J033309.36-280725.8&      &      &        & 	   &       &      &    &      &          &   &           \\
  3& 03:33:20.61& -27:49:10.1&18.47& 0.032&  3.8& J033320.61-274910.3&19.12 & 17.63&  16.68 & 16.12&  0.03 & 32802& G  &      &          &664&$<$1.42e-15\\
  4& 03:32:35.09& -27:55:33.0&18.72& 0.047&  5.2& J033235.09-275533.2&18.58 & 17.26&  16.77 & 16.21&  0.03 & 18675& G  &0.038 &LEX,EN    &247$^d$&1.46e-15$^d$\\
  5& 03:32:29.99& -27:44:04.8&18.75& 0.028&  3.4& J033230.00-274405.0&18.95 & 17.90&  17.50 & 17.10&  0.03 & 42499& G  &0.076 &LEX,EN    &392& 3.05e-15  \\
  6& 03:32:27.01& -27:41:05.1&19.14& 0.077& 13.0& J033227.02-274105.2&19.55 & 19.13&  19.12 & 19.12&  0.98 & 48284& Q  &0.734 &BLAGN     &379& 6.58e-14  \\
  7& 03:31:16.34& -27:57:34.7&19.20& 0.047&  4.6& J033116.35-275735.0&20.15 & 19.47&  19.20 & 18.96&  0.98 & 13879& S  &      &          &   &           \\
  8& 03:32:46.76& -28:08:46.7&19.21& 0.055&  6.5& J033246.76-280846.8&      &      &        & 	   &       &      &    &      &          &   &           \\
  9& 03:33:38.86& -27:40:16.4&19.24& 0.034&  3.0& J033338.87-274016.9&21.53 & 19.78&  19.29 & 18.80&  0.98 &      &    &      &          &   &           \\
 10& 03:32:08.66& -27:47:34.4&19.25& 0.059&  9.2& J033208.68-274734.5&19.78 & 19.07&  19.08 & 18.76&  0.98 & 34357& Q  &0.543 &BLAGN     &305& 7.05e-14  \\
 11& 03:33:16.51& -27:50:39.5&19.39& 0.044&  4.0& J033316.51-275039.7&19.52 & 18.33&  17.78 & 17.27&  0.07 & 28467& G  &      &          &   &           \\
 12& 03:32:45.95& -27:57:45.3&19.60& 0.043&  3.7& J033245.97-275745.6&18.83 & 17.67&  17.26 & 16.89&  0.03 & 14012& G/U&      &          &   &$<$1.06e-15$^e$\\
 13& 03:33:16.07& -27:39:02.8&19.83& 0.040&  3.1& J033316.08-273902.9&23.26 & 20.96&  19.91 & 18.92&  0.98 & 52646& S  &      &          &   &           \\
 14& 03:32:32.00& -28:03:09.9&19.86& 0.100& 11.9& J033232.02-280310.0&19.65 & 19.85&  19.82 & 19.72&  0.98 &  2006& Q  &      &          &398& 3.11e-14  \\
 15& 03:31:54.65& -28:10:35.7&19.88& 0.032&  5.0& ---                &      &      &        & 	   &       &      &    &      &          &   &           \\
 16& 03:33:32.52& -27:38:43.9&19.91& 0.046&  4.0& J033332.53-273844.3&23.03 & 20.77&  19.95 & 19.15&  0.98 & 53315& S  &      &          &   &           \\
 17& 03:31:14.50& -28:10:54.7&19.93& 0.036&  3.3& ---                &      &      &        & 	   &       &      &    &      &          &   &           \\
 18& 03:33:31.37& -27:56:34.2&20.01& 0.089& 10.0& J033331.39-275634.6&20.72 & 20.15&  20.07 & 19.90&  0.98 & 15952& Q  &      &          &723& 3.71e-15  \\
 19& 03:32:26.50& -27:40:35.7&20.02& 0.120& 17.7& J033226.51-274035.7&20.34 & 19.98&  20.07 & 19.94&  0.97 & 49298& Q  &1.031 &BLAGN     &375& 6.16e-15  \\
 20& 03:32:11.84& -28:09:11.0&20.03& 0.040&  6.8& ---                &      &      &        & 	   &       &      &    &      &          &   &           \\
 21& 03:32:44.16& -27:39:42.1&20.03& 0.034&  3.3& J033244.16-273942.3&19.95 & 19.03&  18.48 & 18.10&  0.03 & 51684& G  &      &          &   &$<$4.11e-15$^e$\\
 22& 03:33:28.93& -27:56:41.1&20.05& 0.155& 19.4& J033328.95-275641.4&20.72 & 20.25&  20.12 & 20.12&  0.98 & 15731& Q  &      &          &712& 4.98e-14  \\
 23& 03:32:16.20& -27:39:30.2&20.18& 0.102& 14.5& J033216.21-273930.5&20.74 & 20.36&  20.20 & 19.82&  0.97 & 51593& G/U&1.324 &BLAGN     &345& 7.69e-15  \\
 24& 03:33:09.71& -27:56:14.0&20.21& 0.096& 11.0& J033309.72-275614.3&20.55 & 20.15&  20.11 & 20.05&  0.98 & 16621& Q  &      &          &596& 4.70e-15  \\
 25& 03:31:16.69& -27:43:29.6&20.28& 0.048&  3.5& J033116.70-274329.5&20.34 & 19.43&  19.07 & 18.78&  0.03 & 43124& G  &      &          &   &           \\
 26& 03:32:34.57& -28:03:14.0&20.29& 0.113& 13.6& J033234.59-280314.1&20.91 & 20.30&  20.20 & 20.08&  0.98 &  1821& G/U&      &          &   &           \\
 27& 03:32:38.12& -27:39:44.9&20.48& 0.058&  6.0& J033238.14-273945.0&20.89 & 20.84&  20.43 & 20.41&  0.98 & 50997& Q  &0.837 &BLAGN     &417& 1.47e-14  \\
 28& 03:33:21.09& -27:39:11.8&20.48& 0.078&  7.3& J033321.09-273912.1&20.50 & 20.39&  20.39 & 20.14&  0.83 & 52280& G/U&      &          &670& 1.23e-14  \\
 29& 03:32:37.29& -28:08:47.0&20.50& 0.135& 16.6& J033237.32-280847.3&      &      &        & 	   &       &      &    &      &          &   &           \\
 30& 03:33:12.63& -27:55:51.6&20.50& 0.063&  6.1& J033312.63-275551.9&21.39 & 20.43&  20.38 & 20.08&  0.98 & 17446& G  &      &          &611& 1.17e-14  \\
 31& 03:31:11.38& -27:41:31.9&20.62& 0.102&  9.0& J033111.39-274131.9&20.30 & 20.96&  20.62 & 20.57&  0.98 &      &    &      &          &   &           \\
 32& 03:33:22.79& -27:55:23.8&20.66& 0.150& 18.6& J033322.80-275524.0&20.94 & 21.04&  20.63 & 20.58&  0.98 & 18256& Q  &      &          &678& 4.25e-15  \\
 33& 03:32:53.90& -27:53:54.1&20.74& 0.039&  3.1& J033253.90-275354.3&20.62 & 19.66&  19.13 & 18.75&  0.03 & 21830& G  &      &          &   &$<$7.14e-16$^e$\\
 34& 03:33:26.24& -27:58:29.7&20.79& 0.092& 10.3& J033326.26-275830.0&20.58 & 20.69&  20.65 & 20.57&  0.98 & 11941& Q  &      &          &700& 4.54e-15  \\
 35& 03:32:20.30& -28:02:14.8&20.84& 0.053&  7.4& J033220.34-280214.8&21.03 & 21.17&  20.76 & 20.63&  0.93 &  4050& Q  &      &          &358& 8.71e-15  \\
 36& 03:32:31.78& -28:07:10.4&20.84& 0.045&  3.6& J033231.81-280710.6&      &      &        & 	   &       &      &    &      &          &   &           \\
 37& 03:33:29.22& -27:59:26.7&21.03& 0.075&  7.7& J033329.24-275927.1&21.40 & 21.37&  20.95 & 20.71&  0.92 &  9954& Q/G&      &          &716& 1.37e-14  \\
 38& 03:32:39.09& -27:46:01.8&21.05& 0.051&  4.6& J033239.10-274602.0&21.11 & 21.29&  21.01 & 20.81&  0.98 & 37487& Q  &1.216 &BLAGN     &423& 7.09e-15  \\
 39& 03:32:44.18& -28:10:28.5&21.07& 0.193& 24.7& ---                &      &      &        & 	   &       &      &    &      &          &   &           \\
 40& 03:32:09.44& -27:48:06.8&21.10& 0.058&  6.7& J033209.46-274806.9&22.94 & 21.12&  20.96 & 20.50&  0.98 & 33069& Q  &2.810 &BLAGN     &309& 2.22e-15  \\
 41& 03:31:58.13& -28:02:41.5&21.14& 0.053&  5.8& J033158.14-280241.7&23.02 & 21.57&  20.63 & 19.73&  0.03 &  3111& G/U&      &          &   &           \\
 42& 03:31:20.76& -27:56:48.9&21.14& 0.354& 41.6& J033120.77-275649.2&21.11 & 20.87&  20.88 & 20.77&  0.98 & 15278& Q  &      &          & 21& 1.70e-14  \\
 43& 03:33:20.01& -27:59:12.4&21.22& 0.167& 20.9& J033320.02-275912.7&21.80 & 21.26&  21.26 & 20.92&  0.98 & 10418& G  &      &          &661& 6.46e-15  \\
 44& 03:31:15.04& -27:55:18.6&21.25& 0.047&  3.3& J033115.05-275518.8&21.19 & 21.74&  21.21 & 20.77&  0.91 & 18408& G  &      &          &  7& 3.35e-14  \\
 45& 03:33:07.64& -28:09:51.5&21.27& 0.045&  3.4& ---                &      &      &        &      &       &      &    &      &          &   &           \\
 46& 03:33:06.78& -28:09:14.3&21.42& 0.186& 23.4& ---                &      &      &        & 	   &       &      &    &      &          &   &           \\
 47& 03:32:29.98& -27:45:29.9&21.42& 0.169& 18.4& J033229.99-274530.1&21.51 & 21.41&  21.52 & 21.24&  0.98 & 38551& Q  & 1.218&BLAGN     &391& 1.08e-14  \\
\hline
\end{tabular}
\vfill
\end{minipage}
\end{sidewaystable*}
\newpage
\begin{sidewaystable*}
\begin{minipage}[t][165mm]{\textwidth}
\centering
\begin{tabular}{rccccrccccccrcccrr}
\hline\hline
ID&$\alpha$(J2000)&$\delta$(J2000)&$V$&$\sigma$&$\sigma^*$&EIS-name&$U_{EIS}$&$B_{EIS}$&$V_{EIS}$&$R_{EIS}$&$S_{EIS}$&COMBO&class$^a$&$z$&notes$^b$&XID$^c$&$f_X$(2-8keV)$^c$\\
(1)&(2)&(3)&(4)&(5)&(6)&(7)&(8)&(9)&(10)&(11)&(12)&(13)&(14)&(15)&(16)&(17)&(18)\\
\hline
 48& 03:31:35.43& -28:03:15.8&21.43& 0.051&  3.1& J033135.44-280315.8&22.13 & 21.73&  21.39 & 20.95&  0.90 &  1647& G/U&      &          & 94& 4.48e-15  \\
 49& 03:32:59.86& -27:47:48.2&21.45& 0.205& 22.8& J033259.85-274748.4&22.29 & 22.09&  21.66 & 21.92&  0.98 & 33644& Q  & 2.579&BLAGN     &526& 9.59e-15  \\
 50& 03:33:10.63& -27:57:48.5&21.52& 0.097& 10.6& J033310.64-275748.8&21.68 & 21.47&  21.45 & 21.14&  0.96 & 13332& Q  &      &          &601& 6.41e-15  \\
 51& 03:32:05.22& -28:04:15.3&21.62& 0.071&  5.9& J033205.24-280415.5&22.51 & 22.01&  21.50 & 20.91&  0.98 &      &    &      &          &282& 9.07e-15  \\
 52& 03:31:18.69& -27:41:21.4&21.72& 0.109&  7.4& J033118.71-274121.4&22.18 & 22.75&  21.78 & 21.53&  0.93 & 47501& Q/G&      &          & 12&$<$2.98e-15\\
 53& 03:32:01.18& -28:08:54.8&21.74& 0.157& 16.5& ---                &      &      &        & 	   &       &      &    &      &          &   &           \\
 54& 03:33:30.93& -28:10:55.5&21.75& 0.053&  3.5& ---                &      &      &        & 	   &       &      &    &      &          &   &           \\
 55& 03:31:27.79& -28:00:51.0&21.75& 0.157& 13.0& J033127.80-280051.2&21.54 & 21.78&  21.76 & 21.53&  0.98 &  6817& Q  &      &          & 54& 5.00e-15  \\
 56& 03:33:32.75& -27:49:07.8&21.76& 0.236& 18.9& J033332.77-274908.0&22.23 & 21.59&  22.09 & 21.64&  0.98 & 31085& Q  &      &          &728& 6.63e-15  \\
 57& 03:33:05.31& -27:54:09.1&21.79& 0.191& 14.7& J033305.31-275409.4&22.90 & 22.21&  21.51 & 21.19&  0.54 & 20787& G  &      &SNC       &   &           \\
 58& 03:32:11.83& -28:06:16.3&21.82& 0.086&  6.6& J033211.85-280616.5&21.50 &      &  20.68 & 20.35&  0.03 &      &    & 0.274&SN        &   &           \\
 59& 03:31:29.41& -28:10:27.4&21.87& 0.148& 10.8& ---                &      &      &        & 	   &       &      &    &      &          &   &           \\
 60& 03:32:32.28& -28:03:28.3&21.88& 0.168& 14.1& J033232.30-280328.4&21.85 & 21.89&  21.82 & 21.41&  0.92 &  1257& Q  &      &          &400& 1.88e-14  \\
 61& 03:31:36.25& -28:01:49.7&21.92& 0.466& 36.3& J033136.25-280149.8&22.20 & 21.80&  22.55 & 22.27&  0.97 &  4809& Q  &      &          &100& 5.71e-15  \\
 62& 03:32:59.07& -27:43:39.5&21.92& 0.062&  3.3& J033259.06-274339.8&22.26 & 21.95&  21.48 & 20.99&  0.04 & 42601& Q/G& 0.733&BLAGN     &516& 6.13e-15  \\
 63& 03:32:03.89& -28:10:15.6&21.93& 0.247& 21.9& ---                &      &      &        & 	   &       &      &    &      &          &   &           \\
 64& 03:33:35.56& -27:39:34.7&21.97& 0.127&  7.3& J033335.57-273935.1&23.00 & 22.51&  21.87 & 20.82&  0.06 & 51491& G  &      &          &741& 5.32e-15  \\
 65& 03:31:44.14& -28:05:00.5&21.97& 0.135&  8.7& J033144.17-280500.7&23.68 & 22.64&  22.14 & 21.53&  0.98 &      &    &      &          &148&$<$1.57e-15\\
 66& 03:33:19.04& -28:08:02.5&21.99& 0.368& 33.9& J033319.06-280803.0&      &      &        & 	   &       &      &    &      &          &   &           \\
 67& 03:31:21.45& -28:04:50.5&22.00& 0.144&  9.3& J033121.47-280450.6&22.87 & 21.73&  21.97 & 21.45&  0.89 &      &    &      &          & 26& 6.14e-15  \\
 68& 03:33:06.26& -28:00:55.6&22.02& 0.193& 14.8& J033306.28-280055.8&22.40 & 22.19&  21.72 & 21.73&  0.92 &  6735& Q  &      &          &567& 2.29e-15  \\
 69& 03:31:13.04& -27:50:55.6&22.02& 0.111&  5.5& J033113.05-275055.8&23.67 & 22.55&  21.78 & 20.66&  0.15 &      &    & 0.540&SN        &   &           \\
 70& 03:32:30.19& -28:00:19.9&22.05& 0.138&  9.7& J033230.21-280020.0&22.64 & 22.08&  21.86 & 21.93&  0.90 &  7902& Q  &      &          &393& 6.70e-15  \\
 71& 03:33:19.78& -28:06:27.4&22.06& 0.075&  4.0& J033319.78-280627.8&22.58 &      &  22.01 & 21.09&  0.98 &      &    &      &          &   &           \\
 72& 03:31:35.78& -27:51:34.9&22.11& 0.161&  8.3& J033135.79-275134.9&22.41 & 21.62&  21.95 & 21.65&  0.98 & 25884& Q  &      &          & 96& 1.49e-14  \\
 73& 03:31:51.86& -28:05:54.9&22.24& 0.372& 25.9& J033151.88-280555.1&22.76 &      &  22.46 & 22.25&  0.86 &      &    &      &          &200& 8.71e-15  \\
 74& 03:31:56.86& -28:01:48.7&22.27& 0.276& 18.5& J033156.88-280149.0&22.57 & 22.21&  22.37 & 21.94&  0.83 &  4995& Q  &      &          &235&$<$1.34e-15\\
 75& 03:32:17.14& -27:43:03.3&22.31& 0.091&  4.7& J033217.15-274303.5&23.16 & 22.53&  22.16 & 21.14&  0.85 & 43863& G  & 0.569&BLAGN     &348& 6.29e-15  \\
 76& 03:32:00.37& -27:43:19.7&22.33& 0.192& 12.0& J033200.37-274319.9&22.76 & 22.99&  22.25 & 22.13&  0.98 & 43151& Q  & 1.037&BLAGN     &250& 6.09e-15  \\
 77& 03:31:51.78& -28:00:25.6&22.37& 0.109&  4.7& J033151.80-280025.9&22.88 & 22.58&  22.38 & 22.38&  0.98 &  7671& Q  &      &          &199& 2.25e-15  \\
 78& 03:31:50.95& -27:41:15.9&22.44& 0.160&  8.6& J033150.97-274116.1&23.19 & 22.75&  22.67 & 22.21&  0.95 & 47615& Q  & 0.253&NELG      &192& 6.32e-15  \\
 79& 03:31:17.07& -28:08:20.5&22.45& 0.182&  8.2& ---                &      &      &        & 	   &       &      &    &      &          &   &           \\
 80& 03:31:49.41& -27:46:34.2&22.47& 0.133&  6.6& J033149.42-274634.4&23.26 & 22.72&  22.44 & 22.40&  0.98 & 36120& Q  &      &          &181& 1.26e-15  \\
 81& 03:33:00.78& -27:55:20.7&22.52& 0.529& 28.5& J033300.73-275520.6&22.73 & 23.00&  22.25 & 22.34&  0.37 & 18324& Q/G&      &          &532& 1.42e-14  \\
 82& 03:31:28.61& -28:07:58.6&22.63& 0.330& 14.7& J033128.61-280759.0&      &      &        & 	   &       &      &    &      &          &   &           \\
 83& 03:31:49.54& -27:43:19.4&22.71& 0.178&  7.6& J033149.55-274319.6&23.16 & 22.93&  22.40 & 21.88&  0.97 & 43170& G  & 1.320&q         &184& 7.31e-15  \\
 84& 03:32:10.91& -27:44:15.0&22.75& 0.323& 15.2& J033210.92-274415.2&23.35 & 23.04&  22.90 & 22.42&  0.88 & 41159& Q  & 1.600&BLAGN     &316& 1.27e-14  \\
 85& 03:32:48.57& -28:09:50.5&22.84& 0.167&  5.2& ---                &      &      &        & 	   &       &      &    &      &          &   &           \\
 86& 03:31:40.04& -27:39:17.8&22.85& 0.149&  3.4& J033140.06-273917.8&23.10 & 23.17&  22.93 & 22.94&  0.94 & 51835& Q  &      &          &119&$<$1.59e-15\\
 87& 03:33:22.85& -28:03:13.0&22.85& 0.238&  8.6& J033322.87-280313.2&23.15 & 22.60&  22.71 & 22.32&  0.71 &  1731& Q  &      &          &679&$<$1.58e-15\\
 88& 03:32:38.87& -27:59:18.7&22.86& 0.146&  4.0& J033238.88-275918.9&23.33 & 23.22&  22.58 & 22.04&  0.81 & 10151& G  &      &          &419& 9.43e-15  \\
 89& 03:31:44.21& -28:07:14.1&22.90& 0.219&  6.8& J033144.22-280714.4&      &      &        & 	   &       &      &    &      &          &   &           \\
 90& 03:31:49.95& -28:09:41.6&22.91& 0.355& 14.1& ---                &      &      &        &      &       &      &    &      &          &   &           \\
 91& 03:33:16.08& -28:01:31.3&22.92& 0.135&  3.4& J033316.10-280131.5&22.38 & 22.86&  22.52& 22.21 &  0.60 &  5498& Q/G&      &          &631& 3.94e-15  \\
 92& 03:33:39.28& -28:10:01.4&22.92& 0.150&  4.1& ---                &      &      &       & 	   &       &      &    &      &          &   &           \\
 93& 03:32:49.84& -28:05:14.5&22.93& 0.267&  8.9& J033249.84-280514.8&23.21 &      &  22.64& 22.18 &  0.02 &      &    &      &          &477& 6.28e-15  \\
 94& 03:33:10.19& -27:48:42.0&22.96& 0.208&  5.1& J033310.19-274842.3&24.27 & 23.72&  23.22& 22.77 &  0.87 & 31898& Q/G&      &          &597&$<$1.43e-15\\
\hline
\end{tabular}
\vfill
\end{minipage}
\end{sidewaystable*}
\newpage
\begin{sidewaystable*}
\begin{minipage}[t][165mm]{\textwidth}
\begin{tabular}{rccccrccccccrcccrr}
\hline\hline
ID&$\alpha$(J2000)&$\delta$(J2000)&$V$&$\sigma$&$\sigma^*$&EIS-name&$U_{EIS}$&$B_{EIS}$&$V_{EIS}$&$R_{EIS}$&$S_{EIS}$&COMBO&class$^a$&$z$&notes$^b$&XID$^c$&$f_X$(2-8keV)$^c$\\
(1)&(2)&(3)&(4)&(5)&(6)&(7)&(8)&(9)&(10)&(11)&(12)&(13)&(14)&(15)&(16)&(17)&(18)\\
\hline
 95& 03:33:15.75& -28:08:55.2&22.96& 0.130&  3.1& ---                &      &      &       & 	   &       &      &    &      &          &   &           \\
 96& 03:32:43.24& -27:49:14.1&22.97& 0.332& 13.0& J033243.25-274914.4&22.54 & 22.53&  22.70& 22.65 &  0.95 & 30792& Q  &1.920 &BLAGN     &441& 2.63e-15  \\
 97& 03:33:38.93& -27:42:05.3&23.02& 0.162&  3.2& J033338.94-274205.7&22.63 & 22.29&  21.60& 20.98 &  0.01 &      &    &      &          &   &           \\
 98& 03:32:41.86& -27:52:02.5&23.06& 0.143&  3.6& J033241.86-275202.6&      & 23.79&  23.14& 22.52 &  0.91 & 25042& Q  &3.592 &BLAGN     &435& 2.86e-15  \\
 99& 03:31:14.29& -27:47:07.4&23.13& 0.183&  3.6& J033114.30-274707.5&23.53 & 23.57&  23.06& 22.58 &  0.98 &      &    &      &          &  4&$<$2.54e-15\\
100& 03:32:08.92& -28:09:18.7&23.15& 0.155&  3.5& ---                &      &      &       & 	   &       &      &    &      &          &   &           \\
101& 03:33:21.21& -27:52:19.6&23.16& 0.482& 13.0& J033321.22-275219.6&23.60 & 23.48&  23.59& 23.49 &  0.06 & 24466& G  &      &          &   &           \\
102& 03:32:34.92& -28:09:19.7&23.17& 0.382& 11.4& ---                &      &      &       & 	   &       &      &    &      &          &   &           \\
103& 03:32:32.50& -27:39:02.4&23.25& 0.226&  6.3& J033232.52-273902.6&25.41 & 24.08&  23.33& 22.89 &  0.07 & 52474& G  &      &          &   &$<$1.67e-15$^e$\\
104& 03:33:26.31& -27:48:31.1&23.25& 0.194&  3.3& J033326.32-274831.3&23.75 & 24.06&  23.31& 23.10 &  0.98 & 32254& Q  &      &          &702& 4.99e-15  \\
105& 03:32:52.32& -28:05:38.2&23.27& 0.210&  4.6& J033252.37-280538.4&      &      &  17.64&       &  0.00 &      &    &      &          &   &           \\
106& 03:32:07.32& -28:04:30.7&23.36& 0.165&  3.2& J033207.34-280430.9&      & 23.93&  23.18& 23.09 &  0.96 &      &    &      &          &299&$<$1.80e-15\\
107& 03:33:35.25& -27:51:57.8&23.44& 0.215&  3.2& J033335.27-275158.1&24.28 & 23.85&  23.30& 22.84 &  0.07 & 25213& G  &      &          &   &           \\
108& 03:33:22.96& -27:49:38.1&23.48& 0.367&  7.1& J033322.97-274937.5&23.64 & 23.48&  23.37& 23.69 &  0.15 & 30008& G  &      &          &   &           \\
109& 03:33:39.15& -27:38:51.3&23.53& 0.469&  9.4& J033339.14-273852.1&23.23 & 22.98&  22.47& 21.60 &  0.01 &      &    &      &          &   &           \\
110& 03:32:49.24& -28:09:02.9&23.54& 0.427&  9.9& ---                &      &      &       & 	   &       &      &    &      &          &   &           \\
111& 03:33:35.97& -27:48:03.7&23.59& 0.228&  3.1& J033336.00-274803.9&24.32 & 23.85&  23.18& 22.86 &  0.10 & 33178& G  &      &          &   &           \\
112& 03:31:28.83& -27:52:27.5&23.62& 0.236&  3.3& J033128.84-275227.6&23.95 & 23.36&  22.81& 22.94 &  0.20 & 24036& G  &      &          &   &           \\
113& 03:31:24.15& -27:40:19.0&23.62& 0.242&  3.5& J033124.20-274019.5&23.62 & 23.56&  22.59& 22.38 &  0.00 & 49662& G  &      &          &   &           \\
114& 03:33:37.58& -28:06:00.2&23.62& 0.245&  5.1& J033337.63-280600.6&      &      &  23.60& 22.09 &  0.94 &      &    &      &          &   &           \\
115& 03:32:01.58& -27:43:27.0&23.66& 0.335&  8.5& J033201.59-274327.2&24.86 & 23.58&  23.34& 23.06 &  0.97 & 42882& Q  &2.726 &BLAGN     &260& 4.88e-15  \\
116& 03:33:39.78& -27:56:47.6&23.67& 0.220&  4.1& J033339.81-275647.7&23.92 & 23.21&  22.98& 22.04 &  0.01 &      &    &      &          &   &           \\
117& 03:32:02.43& -28:10:47.3&23.75& 0.205&  3.3& ---                &      &      &       & 	   &       &      &    &      &          &   &           \\
118& 03:32:21.01& -27:40:29.4&23.76& 0.314&  7.3& J033221.02-274030.0&23.69 & 23.48&  23.62& 23.04 &  0.31 & 49352& G  &      &          &   &$<$5.71e-16$^e$\\
119& 03:33:40.03& -28:09:08.3&23.83& 0.207&  3.2& ---                &      &      &       & 	   &       &      &    &      &          &   &           \\
120& 03:33:14.85& -27:57:49.1&23.79& 0.204&  3.1& J033314.86-275749.3&25.35 & 23.70&  23.96& 23.48 &  0.87 & 13244& Q  &      &          &625& 1.05e-15  \\
121& 03:31:47.90& -27:48:31.0&23.85& 0.270&  5.6& J033147.91-274831.2&23.88 & 23.99&  23.42& 23.07 &  0.07 & 32231& G  &0.652 &em        &   &$<$5.13e-16$^e$\\
122& 03:31:59.51& -27:50:21.7&23.89& 0.224&  3.8& J033159.53-275021.6&23.90 & 23.82&  23.66& 23.53 &  0.23 & 28406& G  &      &          &   &$<$5.71e-16$^e$\\
123& 03:31:51.01& -27:39:33.4&23.89& 0.219&  3.6& J033151.04-273933.6&23.99 & 23.71&  23.65& 23.38 &  0.10 & 51258& G  &      &          &   &           \\
124& 03:32:21.64& -27:39:21.5&23.90& 0.266&  5.2& J033221.64-273921.9&24.41 & 23.16&  22.77& 22.16 &  0.22 & 51796& G  &      &          &   &$<$5.49e-16$^e$\\
125& 03:31:47.94& -27:50:45.5&23.91& 0.292&  6.0& J033147.98-275045.5&23.85 & 23.51&  23.78& 23.61 &  0.56 & 27530& G  &1.065 &BLAGN     &170&$<$1.56e-15\\
126& 03:32:35.37& -27:49:20.6&23.92& 0.261&  4.5& J033235.38-274920.8&23.97 & 23.81&  23.78& 23.51 &  0.15 & 30561& G  &0.666 &g         &   &$<$6.00e-16$^e$\\
127& 03:32:42.00& -27:50:51.9&23.93& 0.233&  3.6& J033242.02-275052.0&25.06 & 23.97&  24.06& 23.63 &  0.13 & 27422& G  &      &          &   &$<$5.44e-16$^e$\\
128& 03:31:34.18& -27:38:46.1&23.93& 0.296&  3.8& J033134.16-273846.2&23.62 &      &       &       &       & 52868& G  &      &          &   &           \\
129& 03:33:39.09& -27:52:56.1&23.93& 0.307&  3.8& J033339.09-275256.4&24.19 & 23.79&  22.58& 22.87 &  0.13 &      &    &      &          &   &           \\
130& 03:32:28.65& -27:38:46.5&23.95& 0.250&  4.0& J033228.65-273846.8&24.32 & 24.31&  24.02& 23.57 &  0.10 & 52921& G  &      &          &   &$<$9.66e-16$^e$\\
131& 03:32:35.36& -28:00:41.2&24.00& 0.357&  5.7& J033235.38-280041.4&24.58 & 24.45&  24.07& 23.76 &  0.02 &  7139& Q/G&      &          &408& 2.22e-15  \\
132& 03:33:31.11& -27:59:28.7&24.02& 0.246&  3.8& J033331.13-275929.0&23.96 & 23.97&  23.72& 23.79 &  0.20 &  9765& G  &      &          &   &           \\
\hline
\end{tabular}
~\\
Columns have the following meanings:
(1): object identification No.;
(2) and (3): right ascension $\alpha$ and declination $\delta$ (J2000);
(4): $V$ magnitude of present work (rescaled to $V_{EIS}$);
(5): the standard deviation of the light curve $\sigma$;
(6): the normalised standard deviation $\sigma^*$ (Eq. 3);
(7): EIS-name;
(8-11): $U$, $B$, $V$, $R$  from EIS catalogue (AB system);
(12): stellarity index from EIS catalogue $S_{EIS}$;
(13): COMBO-17 name;
(14): COMBO-17 (SED) class; 
(15):  spectroscopic redshift;
(16): notes on  the nature of the object;
(17): X-ray identification (\# in \citet{lehm05} unless otherwise noted);
(18): X-ray flux (2-8 keV), from \citet{lehm05} unless otherwise noted.

\smallskip
$^a$ COMBO-class: G=galaxy, Q=QSO, S=star, U=unclear \citep{wolf04};
$^b$ BLAGN=Broad Line AGN, LEX=Low Excitation spectrum, NELG=Narrow Emission Line Galaxy \citep{szok04}; q=quasar, em=emission line galaxy \citep{ravi07};
g=galaxy \citep{graz06}; SN=supernova, SNC=supernova candidate \citep{bott08}; EN=extranuclear \citep{horn04,lehm06};
$^c$ From \citet{lehm05} unless otherwise noted;
$^d$ From \citet{giac02};
$^e$ Our estimate

\vfill
\end{minipage}
\end{sidewaystable*}

\end{document}